\documentclass[fleqn,onecolumn,usenatbib]{mnras}  % for review and submission
\title[]{A Variable Polytrope Index Applied to Planet and Material Models} 

\author[]{S.~P.~Weppner,$^1$\thanks{email address: weppnesp@eckerd.edu} J.~P.~McKelvey,$^2$\thanks{deceased July 2014} K.~D.~Thielen $^1$ A.~K.~Zielinski $^1$\\
$^1$Eckerd College; St. Petersburg, FL. 33711, USA \\ $^2$Prof. Emeritus, Clemson University; Clemson, SC. 29634, USA}
\usepackage[utf8]{inputenc}
\usepackage[T1]{fontenc}
\usepackage{ae,aecompl}
\usepackage{graphicx}   % Including figure files
\usepackage{amsmath}    % Advanced maths commands
\usepackage{amssymb}    % Extra maths symbols
\usepackage[usenames, dvipsnames]{color} %For revision purposes
\begin{document}
% revision for mnras
\date{\today}
\maketitle
\begin{abstract}
{We introduce a new approach to a century old assumption
which enhances
not only planetary interior calculations 
but also 
high pressure material physics. 
We show
that the polytropic index is the derivative of the 
bulk modulus with respect to pressure. 
We then
augment the traditional polytrope theory by including a variable polytrope index
{within the confines of the Lane-Emden differential equation.} 
{To investigate the possibilities of this method we 
create} a high quality universal
equation of state, 
transforming the traditional polytrope method to a
tool with the potential for excellent predictive power.
The theoretical foundation of our equation of state is the same elastic observable
which we found equivalent to the polytrope index, the derivative of the
bulk modulus with respect to pressure.
We calculate
the density-pressure of six common materials up to $10^{18}$ Pa, mass-radius relationships
for the same materials, and produce plausible density-radius models for the rocky planets of
our solar system. 
{We argue that the bulk modulus and its derivatives}
{have been under utilized in previous planet formation methods}.
We constrain the material surface observables
for the inner core, outer core, and mantle of planet Earth 
{in
a systematic way including pressure, bulk modulus, and the polytrope index
in the analysis. We believe this variable polytrope 
method has the necessary apparatus 
to be extended further to
gas giants and stars.} As
supplemental material we provide computer code to calculate multi-layered planets.}
\end{abstract}

\section{Introduction}
As more exoplanets are discovered~\citep{cat}, the need for 
a systematic method for planetary models becomes more 
pronounced~\citep{Zapolsky-1969,Stacey2,Valencia2006545,0004-637X-656-1-545,Sotin,Seager:2007ix,Grasset:2009xq,Leconte,
0004-637X-744-1-59,Wagner, Zeng2013,Alibert}.
Our planet formation approach begins with 
the well known polytrope assumption at its foundation.
A polytrope is a simple
structural assumption between a material's pressure
and volume, $PV^n = C$, or similarly it's pressure
and density, $P=C\rho^n$,
where the $C$ is a constant and $n$ is
the polytrope index. 

There have been some attempts to quantify
what the polytrope index represents in the limited context of the
ideal gas law~\citep{Chandrabook2,polybook2}. 
There have been other recent endeavors to import physical meaning
to the polytrope index~\citep{de_Sousa21072011,Christians} but their results were 
limited in
scope.
To our knowledge,
this work exhumes for the first time the physical definition
of the polytrope index in the general case for all materials.

In the body of this article
we will first derive the physical interpretation
of the polytrope index as the pressure derivative
of the bulk modulus and show its kinship to the
Murnaghan equation of state (EOS)~\citep{Murnaghan15091944}.
We then define a new variation of the traditional
Lane-Emden equation 
in Sect.~\ref{S2} and apply 
it in Sect.~\ref{S3}
by fitting the
density-pressure profile of the Earth constructed from seismic data.
Finding this technique limiting we then
permit the polytrope index to vary, modify
the Lane-Emden equation which incorporates it, and apply this
variable index technique to experimental high pressure-density
curves for six common planetary materials up to 
$10^{18}Pa$ in Sect.~\ref{S4}. 
We then, using the same theory, develop a systematic approach
for planetary models and calculate
Mercury, Venus, Earth, 
and Mars (Sect.~\ref{S6}). Analyzing our EOS
relative to previous universal EOSs, 
we spotlight  the functional form of the 
polytrope index.
This survey is done to elucidate  the important
utility this observable has in the {future}
development of more sophisticated
{formation} models.
We finally examine the power of this
technique by scrutinizing the interior of the Earth in Sect.~\ref{S7}
where we constrain, with mixed success, elastic
material observables extrapolated
to the surface of our planet.
We conclude reviewing the potential of this variable polytrope approach.

\section{The polytrope index is a derivative of the Bulk Modulus}\label{S2}

Starting with the definition of the bulk modulus ($B$), the inverse
of the more intuitive compressibility ($K$), we manipulate:
\begin{eqnarray}
\mbox{from }&&B = K^{-1}  = \rho \frac{dP}{d\rho}, \nonumber \\
\mbox{then }&&K^{-1} \frac{d\rho}{dP} = \rho,\nonumber  \\
\mbox{then }&&\frac{dK}{dP} K^{-1} \frac{d\rho}{dP} = \frac{dK}{dP} \rho, 
\end{eqnarray}
and by using the relationship,
$\frac{dK}{dP} K^{-1} = -\frac{dK^{-1}}{dP} K =- \frac{dB}{dP} K$
we are able to progress to
\begin{eqnarray}
\frac{dB}{dP}K\frac{d\rho}{dP}=-\frac{dK}{dP}\rho 
\end{eqnarray}
and multiplying both sides by the awkward $\rho^{\frac{dB}{dP}-1}$ ($\frac{dB}{dP}$ is 
dimensionless)
\begin{eqnarray}
\rho^{\frac{dB}{dP}-1}\frac{dB}{dP}K\frac{d\rho}{dP}
=-\frac{dK}{dP}\rho^{\frac{dB}{dP}}&&\nonumber \\
\mbox{then }\rho^{\frac{dB}{dP}-1}\frac{dB}{dP}K\frac{d\rho}{dP}
+\frac{dK}{dP}\rho^{\frac{dB}{dP}}&=&0 .
\end{eqnarray}
Though this looks complicated we are able to notice a derivative power 
relationship of the form
$\frac{d}{dP}(X^n Y)=X^{n-1}nY \frac{dX}{dP}+\frac{dY}{dP}X^n$
and so
we have the intermediate result 
\begin{equation}
\frac{d}{dP}(\rho^{\frac{dB}{dP}}K)=0 \;\; \mathrm{or} \;\; 
\rho^{\frac{dB}{dP}}K=C^\prime,\label{inter}
\end{equation} 
where $C^\prime$ is a constant.
From the $P=0$ boundary condition we have
\begin{equation}
C^\prime = \rho_0^{({\frac{dB}{dP}})_0}K_0 =\rho_0^{n}K_0 = \rho^{\frac{dB}{dP}}K, \label{inter2}
\end{equation}
where the naught subscript represents
the values of these variables at zero pressure.
This derivation only requires that the exponent, the pressure derivative of the 
bulk modulus, remains constant because
a simple derivative power law was assumed. This relationship, between
density and the compressibility (or bulk modulus), has been shown previously
by
~\cite{Stacey2,Stacey3} in the context of the 
Murnaghan equation of state (EOS) ~\citep{Murnaghan15091944}. This EOS
also 
assumes that
the bulk modulus pressure derivative is constant and thus it reproduces
Eq.~(\ref{inter2}). The derivation 
using the Murnaghan EOS is reproduced in Appendix~A.

This 
result, $\rho^{n}K=C^\prime=\rho_0^{n}K_0$, 
will now be demonstrated as equivalent to a polytrope 
when $n=\frac{dB}{dP}$ is a constant. Because
this derivative is unchanging, it is equivalent to the value at zero pressure,
${(\frac{dB}{dP})}_0 \equiv B^\prime_0$.
The first step in producing the polytrope
is replacing the compressibility with its definition
\begin{equation}
\rho^n\;K = \rho^n\frac{1}{\rho}\frac{d\rho}{dP} = C^\prime.
\end{equation}
This equation is separable with $\rho$ and $P$ so 
multiplying both sides by $dP$ and integrating
\begin{equation}
\int\rho^{n-1}\;d\rho = \int C^\prime\;dP,
\end{equation}
we find
\begin{equation}
\frac{\rho^n}{n} = C^\prime\;P + D, \label{eq:1.1}
\end{equation}
which is in a modified polytrope form allowing
for an additional intercept ($D=\rho_0^n/n$) which can be 
non-zero (solids and liquids
described by the Murnaghan EOS or 
self-bound neutron stars as in \cite{Lattimer:2000nx}). 
In the special case when $D=0$ (if $\rho\approx 0$ when $P=0$) we recover the
normal form of~\cite{Eddington1}:
\begin{equation}
P = \frac{\rho^n}{n\;C^\prime}= C \rho^n,
\end{equation} 
thus using Eq.~(\ref{inter2})
\begin{equation}
\frac{1}{C} = n\;C^\prime= n\rho_0^nK_0 = \frac{n\rho_0^n}{B_0}.\label{inter3}
\end{equation} 
We have successfully recovered
a mathematically equivalent polytrope statement in Eq.~(\ref{inter2}) using 
the bulk modulus and compressibility instead of pressure. 
Our modified polytrope form of Eq.~(\ref{eq:1.1}) is also
equivalent to the Murnaghan EOS ~\citep{Murnaghan15091944},
see Appendix A for a derivation. More importantly we have shown that the
polytrope index is equivalent to the derivative with respect to pressure
of the bulk modulus.
{The physical insight realized by the recognition of this 
elastic observable is significant.
When the traditional polytrope has been successful
($P=C\rho^n$),
most notably in the interior of stars, 
than this derivative
in that star interior is close to constant. So most of the interior
of a main sequence star, which fits closely to a $4/3$ index~\cite{Eddington1},
implies that the derivative with respect to pressure of the bulk modulus is roughly $4/3$.
There has also been polytrope modeling success  with
white dwarfs~\citep{Chandra1931},
brown dwarfs~\citep{Stevenson}, red giants~\citep{Cannon},
and
neutron stars~\citep{Lattimer:2000nx}. Thus the same physical interpretation of the index holds.}
{Also the polytrope constant,$C$ of Eq.~(\ref{inter3}), is now known in 
the interior of the star which puts powerful constraints
on the formation constituents. It requires
that 
\begin{equation}
\frac{B_i}{\rho^n_i} = nC = \mbox{constant},
\end{equation}
where
$B_i$ and $\rho_i$ are the bulk modulus and density, at any spot in the interior
of the star when $n \equiv \frac{dB}{dP} =\mbox{constant}$.}

\section{Developing the Fixed Polytrope Technique}\label{S3}
Now on to deriving the gravitational differential equation
for large spheres.
This derivation is similar as to those traditional derivations
found in ~\cite{Chandrabook2} 
and ~\cite{polybook2} except now we 
use the modified form of our polytrope, 
$\rho^{n}K=\rho_0^{n}K_0$, from Eq.~(\ref{inter2}), instead of the 
traditional Eddington form, $P = C\rho^n$. One advantage being 
that Eq.~(\ref{inter}) is more general and holds even when
the density does not approach zero in the limit of zero pressure,
another is that we replace the pressure as our independent variable 
with elastic constants.

If $P$ (pressure) and $g$ (gravity)
are functions of the radial distance 
$r$, as measured from the core,
one may start by considering the equilibrium of radial force components 
on a concentric spherical shell element of interior radius $r$ and exterior 
radius $r + dr$, within which differential changes $dr$, $dP$, $d\rho$ and 
$dg$ alter their respective variables.  Writing the forces as pressure times 
area and mass times gravitational acceleration and equating radial force 
components to zero, with the neglect of second order differentials, one 
obtains a differential equation for the pressure of the form
\begin{equation}
\frac{dP}{dr}=-\rho(r)g(r)\label{le1}
\end{equation}
This equation expresses the Newtonian equilibrium of hydrostatic forces. 
The density is affected by pressure through a compressibility relation of the form
\begin{equation}
K(\rho)=-\frac{1}{V}\frac{dV}{dP}=\frac{1}{\rho}\frac{d\rho}{dP}\label{le2}
\end{equation}
In writing these expressions it is to be noted that in a volume element whose 
mass is conserved, $dm = 0$, and thus
$0=d(\rho V)=\rho dV+d\rho$
and thus
$\frac{dV}{V}=-\frac{d\rho}{\rho}$.
From Eq.~(\ref{le2}) it is easily seen that 
$\rho K(\rho) = d\rho/dP$, while $d\rho/dr = (d\rho/dP)(dP/dr)$.  
Therefore, Eq.~(\ref{le1}) can be written as 
\begin{equation}
\frac{d\rho}{dr}=-\rho ^{2}(r)K(\rho)g(r). \label{le4}
\end{equation}
This expression depends upon an understanding of how 
the compressibility $K$ varies as a function of density. 
First one must observe from Eq.~(\ref{le4})  that
\begin{equation}
g(r)=-\frac{1}{\rho ^{2}K(\rho)}\frac{d\rho}{dr}.\label{le5}
\end{equation}
Also, the mass $m(r)$ within a sphere of radius $r$ is given by 
$m(r)=\int_0^r 4 \pi r^{2} \rho (r)\mathrm{d}r.$
It is now clear that $g(r)$ can also be written as 
\begin{equation}
g(r)=\frac{Gm(r)}{r^{2}}=\frac{G}{r^{2}}\int_0^r 4 \pi r^{2} \rho (r)\mathrm{d}r, \label{le7}
\end{equation}
where $G$ is the Newtonian universal gravitation constant. Differentiating this expression, 
one now obtains 
\begin{equation}
\frac{dg}{dr}=G\left[4 \pi \rho (r)-\frac{2}{r^{3}} 
\int_0^r4 \pi r^{2} \rho (r) \mathrm{d}r\right].
\end{equation}
Using this result along with Eq.~(\ref{le5}), we may now write
\begin{equation}
\frac{d}{dr}\left[-\frac{1}{\rho^{2}K(\rho)}\frac{d\rho}{dr}\right]=
G\left[4\pi\rho(r)-\frac{2}{r^{3}} \int_0^r4 \pi r^{2} \rho (r) \mathrm{d}r\right].
\end{equation}
In this expression, the second term on the right side can be identified with the help of 
Eq.~(\ref{le7}), as  $2G/r^3\int_0^r 4 \pi r^{2} \rho (r) \mathrm{d}r$
$ = 2g(r)/r$, 
which with the further aid of Eq.~(\ref{le5}), leads finally to
\begin{equation}
\frac{d}{dr}\left[-\frac{1}{\rho^{2}K(\rho)}\frac{d\rho}{dr}\right]
-\frac{2}{r\rho^{2}K(\rho)}\frac{d\rho}{dr}-4\pi G \rho (r)=0. \label{le20}
\end{equation}
This can be put in a more explicit form by performing the indicated differentiation with 
respect to $r$ in the leading term.  In so doing it is convenient 
to make a substitution of the form
$\nu(\rho)=\rho^{2}K(\rho)$, 
and using this relationship:
$\frac{d\nu}{dr}=\frac{d\nu}{d\rho} . \frac{d\rho}{dr}$.
The differentiation with respect to $r$ will clearly yield a nonlinear term containing a 
factor of the squared derivative $(d\rho/dr)^2$.  The algebra is straightforward though 
tedious, but finally leads to a nonlinear differential equation of the form
\begin{equation}
\frac{d^{2}\rho}{dr^{2}} - \left[\frac{2}{\rho} + 
\frac{K'}{K}\right]\left(\frac{d\rho}{dr}\right)^{2} + 
\frac{2}{r}\frac{d\rho}{dr} + 4\pi G \rho ^{3}K(\rho)=0 \label{le12},
\end{equation}
where $K'$ is the derivative of the compressibility
with respect to density.
One can also alternatively use Eqs.~(\ref{le2},\ref{le20}) to remove
compressibility for pressure and get a version
equivalent to the original Lane-Emden form used by Eddington,
\begin{equation}
\frac{d^{2}P}{dr^{2}} + \left[\frac{2}{r} - \frac{1}{\rho}\frac{d\rho}{dr}
\right]\frac{dP}{dr} + 4\pi G \rho ^{2}=0. \label{le15}
\end{equation}
This traditional version is mathematically simpler than the version derived above
but it cannot use our modified polytrope assumption,
$\rho_0^{n} K_0 = \rho^n K$, so we keep the more complicated Eq.~(\ref{le12}).

Now assuming the density-compressibility relationship of 
our modified polytrope
\begin{equation}
\rho_0^{n} K_0 = \rho^n K \quad \mathrm{or} 
\quad K=\frac{\rho_0 ^{n}K_0}{\rho^{n}}, \label{le13}
\end{equation}
Where $\rho_0$ and $K_0$ are the density and compressibility of a material at 
vacuum pressure, and then differentiating to get 
\begin{equation}
K'= \frac{-n\rho_0^n K_0}{\rho^{n+1}}
\end{equation}
We can reconstruct Eq.~(\ref{le12}) as
\begin{equation}
\frac{d^{2} \rho}{dr^{2}}-\left[\frac{2-n}{\rho}\right]\left(\frac{d\rho}{dr}\right)^{2} + 
\frac{2}{r}\frac{d\rho}{dr}+4\pi G \rho_0 ^{n}K_0\rho ^{3-n} = 0 .\label{le14}
\end{equation}
A difference between our version of the
Lane-Emden equation, Eq.~(\ref{le14}), and the traditional 
equation, Eq.~(\ref{le15}),
is that ours already has assumed the polytrope assumption, Eq.~(\ref{le13}),
 while Eq.~(\ref{le15}) has not.
If the polytrope assumption of Eq.~(\ref{eq:1.1})
is used to remove pressure dependence in Eq.~(\ref{le15}) than
it will be mathematically congruent with our Eq.~(\ref{le14}).
This equivalence will be true for both zero and non-zero $D$.

\begin{figure*}
\includegraphics[width=120mm]{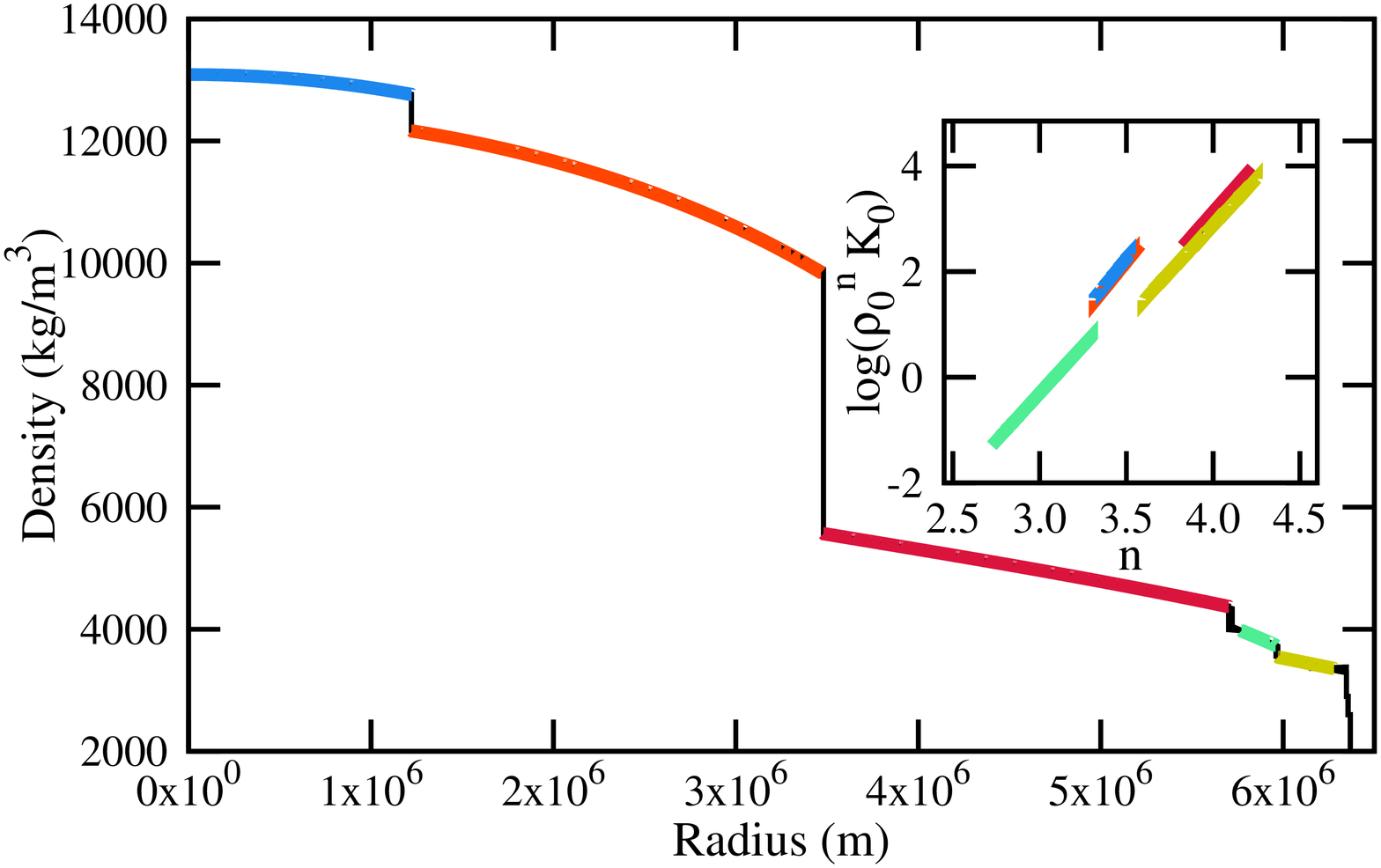}
\label{fig1}
\caption{(color online) In black,
significantly buried under the theoretical calculations, is the
Preliminary Reference Earth Model~\citep{Dziewonski81} (PREM) and the
newer Reference
Earth Model~\citep{kustowski08} (REF). The colored lines are a sample calculation of ours
for inner core (blue), outer core (orange), lower mantle (red),
transition (light green) and upper mantle (yellow). There is excellent
(less than 1\%) disagreement between theory and experiment. The inset
depicts the linear range of the polytrope index that 
share excellent agreement with experimental data using as a legend
the same
colors as the main plot.
}
\end{figure*}

In Fig.~1 we show a density profile of planet Earth 
from the seismic analysis of the {\it Preliminary 
Reference Earth Model}~\citep{Dziewonski81} (PREM)
and the newer {\it Reference Earth Model}~\citep{kustowski08} (REF) giving a profile of Earth's 
inner core, outer core, and three stages of the mantle. Most of the seismic data is obscured
by
our calculation of the solution to Eq.~(\ref{le14}) 
using our fixed polytrope assumption.
Our calculation returned excellent results for the density profile and the gravitational
and pressure profiles as well (not shown) using a variety of different 
vacuum densities, $\rho_0$, bulk moduli, $B_0$, and polytrope indices which are 
equal to the bulk modulus pressure derivative ($\frac{dB_0}{dP}\equiv B^\prime_0\equiv n$). 
The inner most boundary had its density set at the value from the experimental data set, 
the first derivative initial value was constrained 
using Eq.~(\ref{le4}) and Eq.~(\ref{le13}).
The poorest fit was for the liquid outer core but results were still
well under 1\% error.
However the predictive power of this fixed index analysis is weak. With an ideal theory,
one would 
assume that by choosing vacuum values which best fit the experimental seismic data
one could elicit more detail about the composite materials comprising the interior Earth. 
Unfortunately
a plethora of values for $\rho_0, B_0$, and $B^\prime_0$ give excellent
simultaneous fits to both the density-radius and the density-pressure
profile.
The inset of Fig. 1 details this lack of clarity. Each theoretical result
which provided a good agreement to the subset seismic data
of the interior Earth is provided by a point 
on a line of the form $\log_{10}(\rho_0^nK_0) = m\;n + b$. 
The polytropic index
$n$ worked well over a large range as long as 
$\rho_0$ and $K_0$ were constrained to the displayed line. 
By using an EOS which fixes this polytrope observable means that the best fits
will reflect only the best average values of $(\frac{dB}{dP})$ for that span
of the interior of the Earth. 
{For example the inner core
is best reproduced with a fixed index range of $3.35 < n < 3.55$ but it is 
know that at the surface
the same materials, rich in iron, will have $n>4.5$~\citep{Seager:2007ix,Xian10,garai2011pvt,GRL:GRL29665}.}
Having a fixed polytropic index
for planets does not represent the physical reality adequately, despite the fact
that the pressure-density and density-radius profiles were described satisfactorily.
{Unlike the solar models which do have a near constant
pressure derivative of the bulk modulus,
the planet density-radius models, due to their relatively high compressibility, 
are rather insensitive to the 
pressure derivative of the bulk modulus (our newly discovered polytrope index).}
We will delve further into the analysis of the seismic data for Earth in Sect.~\ref{S7},
but next we advance a variable polytrope procedure which will give our model more physical
insight.

\section{Making the Polytropic Index a Variable}\label{S4}
The ansatz
that the polytropic index is constant is constraining and unrealistic.
Eddington inherently
recognized this with star formation
and attempted a variable index analysis~(\citeyear{Eddington2}).
It is well known that for all materials
the derivative of the bulk modulus with respect to pressure does change
($\frac{\partial^2B}{\partial P^2} \ne 0)$~\citep{JGRB:JGRB15234,1.2822458,Singh-2010}.
We show later that the bulk modulus pressure derivative can 
best be approximated as a constant 
only at pressures below $10^{9}$ Pa (small rocky planets or asteroids) or 
above $10^{14}$Pa (interiors of stars and very massive planets). Most
planets have a pressure at their core which is between these extremes, including our Earth
and all the planets in our solar system.
Likewise the Murnaghan EOS was quickly
deemed unsatisfactorily because of its constant derivative of the bulk
modulus with respect to pressure so Birch proposed an
extension~\citep{Birch}. 
A more 
realistic approach would allow the index to vary while at the same time retaining
the definition for the polytrope index,
$n=\frac{dB}{dP}$. 

\subsection{Modifying the Polytrope}
So first assuming a modified
polytrope with a variable index we must allow for the contingent that the polytrope
expression, derived in Sect.~\ref{S2}, is no longer constant
\begin{equation}
\rho^{n(\rho)} K = \rho_0^{n(\rho)} K_0 h(\rho), \label{m1}
\end{equation}
where we have introduced a new function, $h(\rho)$ which is a measure of the
non-conservation of the fixed expression or a weighting function. In the limit of the fixed case
$h(\rho) = 1$.

By taking the natural log of this equation and then taking the pressure derivative 
\begin{equation}
\frac{d}{dP}\left (n(\rho)\ln \frac{\rho}{\rho_0} +\ln \frac{K}{K_0}\right ) = \frac{d}{dP} \ln h(\rho),
\end{equation}
and progressing by realizing that $B=K^{-1}$, $\frac{d}{dP}\ln \rho = K$, and
thus $\frac{d}{dP}\ln K =-\frac{dB/dP}{B}$ 
\begin{equation}
\frac{dn(\rho)}{dP}\ln\frac{\rho}{\rho_0}+ \frac{n(\rho)}{B}-\frac{dB/dP}{B} 
=  \frac{d}{dP} \ln h(\rho),
\end{equation} 
We chose to maintain $n = \frac{dB}{dP}$, thus two terms cancel and we have
\begin{equation}
\frac{dn(\rho)}{dP}\ln\frac{\rho}{\rho_0} =  \frac{d}{dP} \ln h(\rho). \label{m2}
\end{equation}
This is a simple prescription on how use a modified 
polytrope with a variable index equivalent to
the pressure derivative of the bulk modulus.
We now have a new replacement for compressibility, using Eq.~(\ref{m1}) we have
\begin{equation}
K = \frac{\rho_0^{n(\rho)} K_0 h(\rho)}{\rho^{n(\rho)}}, \label{m3}
\end{equation}
which is analogous to the fixed $n$ version, Eq.~(\ref{le13}), with the 
addition of a weighting function, $h(\rho)$.

\subsection{Changes to the Lane-Emden Equation}
The new modified
Lane-Emden equation, which now includes a variable $n$. Starting 
with
\begin{equation}
\frac{d^{2}\rho}{dr^{2}} - \left[\frac{2}{\rho} + 
\frac{K'}{K}\right]\left(\frac{d\rho}{dr}\right)^{2} + 
\frac{2}{r}\frac{d\rho}{dr} + 4\pi G \rho ^{3}K(\rho)=0 \nonumber,
\end{equation}
it is trivial to remove $K$ in terms of $\rho$ using Eq.~(\ref{m3}). We also need to
find $K' = \frac{dK}{d\rho}$, which is a larger undertaking since $n$ is
no longer constant and needs to be treated with care. Starting again with Eq.~(\ref{m1})
we take the logarithm of both sides
\begin{eqnarray}
\frac{d}{d\rho}\log K &=& \frac{d}{d\rho}(\log h 
+ \log\rho_0^{n}K_0 - n \log \rho) \nonumber \\
&=& \frac{d}{d\rho}\log h -\frac{dn}{d\rho}\log \rho 
- n \frac{d}{d\rho}\log \rho+\frac{dn}{d\rho}\log \rho_0, 
\end{eqnarray}
recognizing that we can use earlier relationships if we assume a natural log
\begin{equation}
\frac{dK}{d\rho}= K \left(\frac{d\ln h}{d\rho} 
+\frac{dn}{d\rho}\ln \frac{\rho_0}{\rho} - \frac{n}{\rho}\right)
\end{equation}
and then using Eq.~(\ref{m2}) and the definition 
$K=\frac{1}{\rho}\frac{d\rho}{dP}$ a cancellation occurs and 
we are left with the same result 
as with a fixed $n$
\begin{equation}
\frac{dK}{d\rho}=K^\prime=K(- \frac{n}{\rho}),
\end{equation}
except now $n$ is no longer a constant but a density dependent 
function.
Thus the only symbolic change to the Lane-Emden equation is in the final term
\begin{equation}
\frac{d^{2} \rho}{dr^{2}}
-\left[\frac{2-n(\rho)}{\rho}\right]\left(\frac{d\rho}{dr}\right)^{2} + 
\frac{2}{r}\frac{d\rho}{dr}+4\pi G \,h(\rho)\, \rho_0 ^{n(\rho)}K_0\rho ^{3-n(\rho)} = 0,\label{eq7}
\end{equation}
where the function $h(\rho)$ is introduced 
and $n$, the polytrope index and equivalent to the derivative of the 
bulk modulus with respect to pressure, is also assumed a function of density as 
expected.
This derivation is independent of the EOS function chosen for $n(\rho)$ which,
by Eq.~(\ref{m2}), is also intrinsically linked to the weighting function $h(\rho)$.

\subsection{Our Equation of State}
To have a working hypothesis for a dynamic polytrope index EOS we examined previous
equations of state~\citep{0953-8984-17-39-007,0953-8984-18-46-015}
and pressure-density relationships~\citep{Seager:2007ix,0004-637X-744-1-59}
for an empirical function form for the changing polytrope index.
To be consistent we developed our own EOS which uses the 
informed polytrope index 
as the function of reference in place of 
{the more common pressure EOS
like the Birch-Murnaghan~\citep{Birch,earth} or 
Vinet~\citep{JGRB:JGRB6242,0953-8984-1-11-002,earth}.} 
We found that 
\begin{equation}
n(\rho) = \frac{dB}{dP}=A_0\left(\frac{\rho_0}{\rho}\right)^{A_1} + A_2 \label{eos1}
\end{equation}
works well (more detail on how this empirical 
equation was developed is proffered in Sect.~\ref{S6}). 
It is a respectable EOS for pressures up to  
the $10^{13}$ Pa range.
The resulting functional form for the polytrope index is also exceedingly simple.
All other major relations to this polytrope index
can be found analytically by integration or 
differentiation, using 
fundamental relations found in Appendix \ref{oureos}.
Specifically applying these relations to our 
EOS as expressed in Eq.~(\ref{eos1}) we find for pressure
\begin{equation}
P= \frac{B_0e^{\frac{A_0}{A_1}}}{A_1}
\left [\left(\frac{\rho}{\rho_0}\right)^{A_2}E_n\left(\frac{A_1+A_2}{A_1}
,\frac{A_0}{A_1}\left(\frac{\rho_0}{\rho}\right)^{A_1}\right)  
- E_n\left(\frac{A_1+A_2}{A_1},\frac{A_0}{A_1}\right)\right ]. \label{ourpressure}
\end{equation}
This pressure equation
contains a special function, 
the generalized exponential integral ($E_n$) which is defined as
\begin{equation}
E_n(n,x) = \int_1^\infty \frac{e^{-xt}}{t^n}dt.
\end{equation}

The parameters
$A_0, A_1,A_2$ are dimensionless, connected to experiment, and 
well behaved, always greater than zero and less than ten.
They are connected to the extremes of the polytrope index: 
the infinite pressure 
derivative of the bulk modulus with respect to pressure,$B^\prime_\infty = n_\infty$ and
the zero pressure asymptote of the same observable, $B^\prime_0 = n_0$.
They were fit as follows:
$A_2 = B^\prime_\infty, A_0 = B_0^\prime - A_2$, and 
$A_1 = \frac{-B_0B_0^{\prime\prime}}{B_0^\prime - A_2}$.
In this work we used the parameter settings in a predictive method, 
setting $B_0$, and $B_0^\prime$ close to experimental
results, choosing a near fixed, but yet to be determined,
infinite pressure derivative
($1.75 < B^\prime_\infty < 2.1$),
and assuming a universal ratio for $B_0^{\prime\prime}$
by setting
\begin{equation} 
A_1 = \frac{B_0^\prime}{B_0^\prime-A_2}\equiv \frac{B_0^\prime}{A_0}, \label{roy2}
\end{equation}
which we borrowed from \cite{0953-8984-18-46-015}. Thus
for any given material there are only three parameters ($\rho_0,B_0,$ and
$B^\prime_0$). We
will show our EOS compared to experiment in subsection~\ref{S5}
 but first we require that it be continuous with a 
higher pressure theory to extend its validity.

\subsection{Pressures above Ten Tera-Pascal}\label{hp}
Inspired by~\cite{Seager:2007ix} we went higher in 
pressure, to $10^{18}$ Pa, by following their technique 
to match at some critical density 
our empirical EOS to the 
Thomas-Fermi-Dirac theory~\citep{Salpeter:1967zz} which treats at
extreme pressures the quantum mechanic Fermi-Dirac gas whose pressure we
approximated as
\begin{equation}
P(\rho) = P_0 +F_3\rho
+F_2\rho^{4/3}+F_1\rho^{5/3} \label{tfd2}
\end{equation}  where
\begin{eqnarray}
F_1 &=& \frac{5.16\times 10^{12} Pa}{(2690\frac{A}{Z})^{5/3}} \nonumber \\
F_2 &=& \frac{5.16\times 10^{12} Pa}{(2690\frac{A}{Z})^{4/3}}
(.40726Z^{2/3}+.20732)  \nonumber \\
F_3 &=& \frac{5.16\times 10^{12} Pa}{(2690\frac{A}{Z})^{3/3}}(.01407),\label{tfd4}
\end{eqnarray}
the variable $P_0$ of Eq.~(\ref{tfd2}) 
is a small, relatively low pressure, addition to keep pressure continuity across
the matching boundary and
will be determined last. If the material is not monatomic we use a 
weighted average for $A$ and $Z$ (atomic and proton number respectively).

Following the common definition used between the pressure and the bulk 
modulus, $B = \rho\frac{dP}{d\rho}$,
we have for this high pressure theory
\begin{equation}
B = (5/3) F_1 \rho^{5/3}- (4/3) F_2 \rho^{4/3} - (3/3) F_3 \rho ^{3/3}.\label{tfd5}
\end{equation}
Likewise 
$n$, our polytrope index equals 
\begin{equation}
\frac{dB}{dP} = n=\frac{\rho}{B}\frac{dB}{d\rho},\label{eq1.4}
\end{equation}
thus
\begin{equation}
n=\frac{(5/3)^2 F_1 - (4/3)^2 F_2 \rho^{-1/3} - F_3 \rho ^{-2/3}}
{(5/3) F_1 - (4/3) F_2 \rho^{-1/3} -  F_3 \rho ^{-2/3}}.\label{tfd1}
\end{equation}
A key aspect of this method is to choose the critical density in which to switch from 
$n(\rho) = A_0(\frac{\rho_0}{\rho})^{A_1} + A_2$, to the $n$ of Eq.~(\ref{tfd1}).
We found that by requiring at the boundary
that the bulk modulus, $B$, 
was continuous and using the secant method to 
find this critical density was all that
was required.
We also found that at this critical density
$n$, the derivative of the bulk modulus, also was continuous. 
By adding a relatively small constant to 
pressure ($P_0$ of Eq.~(\ref{tfd2})) we can also set 
the pressures equivalent at the critical density boundary.
The reason we get  a match between both $B$ and $n$ at the boundary is because these
two variables are not independent. Recalling Eq.~(\ref{eq1.4}),
$\frac{dB}{dp} = n=\frac{\rho}{B}\frac{dB}{d\rho}$, we see that if $\rho$ and $B$ are 
attuned we get the two derivatives, $n = \frac{dB}{dP}$ and $\frac{dB}{d\rho}$ as bonus.
With this extension to a high pressure EOS, we now feel comfortable in fitting up to $10^{18}$ Pa
which are core pressures typical of small stars.

Pragmatically we assume informed values for a material's $\rho_0,B_0$, and $B^\prime_0$.
We then make initial approximations for $A_2\approx 2$. We then solve for $A_0,$ and 
$A_1$ of Eq.~(\ref{eos1}). Then 
a secant algorithm adjusts $A_2$ and then determines $A_0,$ and $A_1$ so at the critical
density, $\rho_c$, the low pressure EOS becomes continuous with the higher 
pressure Thomas-Fermi-Dirac EOS (TFD)
of Eq.~(\ref{tfd5}) by finding at what critical density $B_{ours}-B_{TFD}=0$. We then
ensure that $n_{ours}-n_{TFD}=0$ and then we lastly solve for $P_0$ of Eq.~(\ref{tfd2}) to establish
continuity for pressure, bulk modulus, and the pressure derivative of the bulk modulus  
across the critical boundary. As supplemental material we provide an example code 
to build a multi-layered planet
which
has this methodology within (see Sect. \ref{S6}).
 
\subsection{Comparison with experimental results}\label{S5}

\begin{figure*}
\includegraphics[width=120mm]{fig2.eps}\label{fig2}
\caption{(color online) We describe
six materials with our EOS with a solid black line. In the bottom panel an older
EOS, the Vinet~\protect\citep{JGRB:JGRB6242,0953-8984-1-11-002,earth}, is represented by a green dashed line
(parameters for 
H$_2$~\protect\citep{Loubeyre},parameters for He~\protect\citep{PhysRevLett.71.2272},parameters for 
H$_2$O~\protect\citep{high}, we
modified $B^\prime =7.25$, well with in range to give a better fit).
In the top panel an older
EOS, the Birch-Murnaghan~\protect\citep{Birch,earth},
is also represented by a green dashed line
(parameters for MgO~\protect\citep{Duffy95}, parameters for SiO$_2$~\protect\citep{Driver10,Ping13}), we took
an average value between  the two references for $B = 305$ GPa and $B^\prime =5.0$). The iron
green dashed line
is represented by the Vinet with parameters given by~\protect\cite{GRL:GRL13615}.
More sophisticated first principle EOS (density functional theories)
are a red
or orange dashed line (H$_2$-DFT~\protect\citep{Geng12},He-GGA~\protect\citep{Khai08},
H$_2$O-QMD (red)~\protect\citep{French09} and H$_2$O-(orange)~\protect\citep{Hermann29122011},
MgO-GGA-B1 (red) and 
qMS-Q ff (orange)~\protect\citep{Strach99},SiO$_2$-DFT~\protect\citep{Driver10},Fe-EXAFS~\protect\citep{Ping13}).
Experimental data is expressed in purple, blue, and turquoise triangles (H$_2$-purple~\protect\citep{Zha93},
blue~\protect\citep{Geng12},He-purple~\protect\citep{Cazorla08},H$_2$O-purple~\protect\citep{sugimura08},
MgO-purple,blue,turquoise~\protect\citep{Duffy95}
SiO$_2$-purple~\protect\citep{Panero03b},blue~\protect\citep{Xun10},Fe-purple, blue, turquoise~\protect\citep{Xian10}).
The MgO curve is offset
for aesthetic reasons, so that the calculation does not run into the calculation
for SiO$_2$. The inset shows the
changing value of the
polytrope index as the pressure changes in our EOS. All the parameters used for our calculation in this
plot can be found in Appendix~C.}
\end{figure*}

In Fig. 2 we plot density versus pressure for 
six common materials that make up
planet interiors in our solar system ranging from molecular hydrogen
to atomic iron. The materials are either solids or liquids
throughout the calculation and are also low temperature isothermal calculations
(cold curves, 0K - 300K).
We choose to examine an intermediate pressure range, $10^9$ Pa to $10^{12}$ Pa, 
because it is
challenging; it is at the extreme of our experimental capabilities and 
the interiors of our solar system planets are in this range. The
solid black line of the inset 
is our EOS which was input into the Lane-Emden equation of Eq.~(\ref{eq7})
through 
the polytropic index (as a check of our work we used Eq.~(\ref{ourpressure})). 
At these lower pressures we used the EOS described 
with Eq.~(\ref{eos1}), the
values of the parameters can be also found in Appendix \ref{param}.
The filled triangles
 of various
colors are experimental data, the dashed green line is a common empirical theory from the twentieth century
(Birch-Murnaghan~\citep{Birch} or Vinet~\citep{JGRB:JGRB6242,0953-8984-1-11-002,earth} 
which begins to fail in the $10^{11}$ Pascal range, the dotted red
and orange lines are density functional theories and/or quantum Monte Carlo 
theories from the twenty-first century.
Our theory has the correct general trends, it fits theories and experimental data well 
considering that it has no adjustable parameters. Our method was that 
we first chose average experimental
values for our material vacuum values of $\rho_0,B_0,$ and $n_0$. 
In many cases the literature values varied widely and so some
prudence was used to determine the selected values. Our parameter $A_2$ 
was chosen as the critical boundary
value for $n=\frac{dB}{dP}$ when the EOS transitioned from our own to the 
Thomas-Fermi-Dirac EOS~\citep{Salpeter:1967zz} ($1.75 < A_2 < 2.1$) 
and we fixed out remaining parameters by constraining $A_0+A_2 = n_0$ and
$A_0A_1 = n_0$. All the parameters used to calculate the pressure-density
profiles of Fig.~2 can be found in Appendix C.

This success is proof of
concept, that it is possible to develop an experimentally informed 
EOS which uses a polytrope function form, equivalent to $\frac{dB}{dP}$,
which can be naturally and consistently input into the dynamic index Lane-Emden equation.
{It rivals the popular universal Birch-Murnaghan~\citep{Birch,earth} and
Vinet~\citep{JGRB:JGRB6242,0953-8984-1-11-002,earth} in predictive potential as it mimics
well the} 
high pressure first-principle theories~\citep{French09,Geng12,Khai08,Hermann29122011,Driver10,Ping13}. 
The inset of Fig. 2 plots the polytrope index as a 
function of the same pressure range for the six materials.
The polytrope index, $n=\frac{dB}{dP}$ does change significantly over this pressure range, 
thus to keep the polytrope index fixed for planets
in our solar system would be worse than approximate.

\begin{figure*}
\includegraphics[width=120mm]{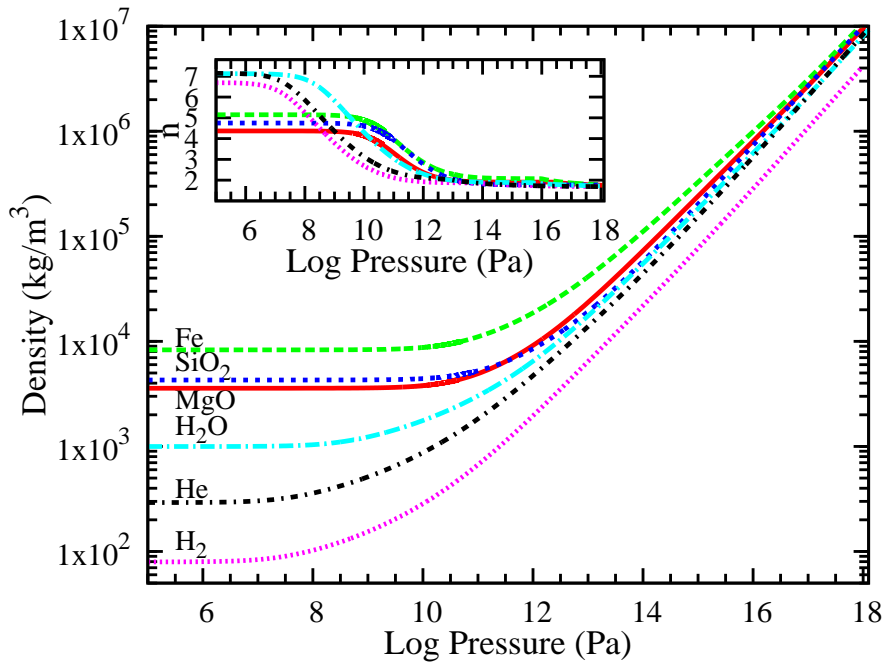}\label{fig3}
\caption{(color online) Our EOS calculation
shown over a large pressure range similar to Fig.~2.
The inset also has the same form as the inset of Fig.~2, the
change in the polytrope index as a function of the same pressure range as the
main plot.}
\end{figure*}

Figure 3 shows  our calculation over a wider 
pressure range, with a similar inset as Fig. 2. Note the 
general trends: up to $10^7$ Pa the pressure density profile is a nearly incompressible
horizontal
line, the polytrope index, equivalent to the pressure 
derivative of the bulk modulus, wanders little from 
from its zero pressure value. The most dynamic range of change for the polytrope index is from 
$10^8$ Pa to $10^{14}$ Pa as was illustrated in Fig. 2. 
As the pressure increases, our theory must switch from
the low pressure EOS of Eq.~(\ref{eos1}) to the technique of Thomas-Fermi-Dirac of
Sect.~\ref{hp}.
The polytrope index again becomes nearly constant after $10^{14}$ Pa as each material
begins its slow asymptotic drive towards 
the Thomas-Fermi-Dirac value of $5/3$~\citep{Salpeter:1967zz}. The parameters 
used for the materials in this figure can be found in Appendix \ref{param}.
{A word of caution; since this is a simple universal EOS it 
will not contain any insight about quantum
phase transitions which will take place in many
of these materials above $10^{11}$ Pascals.}

{As one approaches solar pressure range, as we do in Fig. 3 ($> 10^{15}Pa$), 
it is interesting to note that the polytrope index is again near constant.
Was Eddington's choice of the polytrope index set to $4/3$ for main
sequence stars~\citep{Eddington1} thus fortuitous? Assume, in a main sequence star,
the most significant interactions under extreme
pressure and temperature are gravity, electron repulsion, and heat flow. Eddington assumed
only heat flow
and gravity but completely neglected the significant electron repulsion in
a dense solar gas yet still
calculated satisfactory results. If we examine the
derivative of the bulk modulus with respect to pressure, our newly revealed polytrope index,
we discover why. For large massive objects like suns their partial bulk moduli with
respect to pressure have similar values, greater than one and less than two ($\approx 4/3$ for
heat flow (Eddington) and $\approx 5/3$ for electron repulsion if the  Thomas-Fermi-Dirac scheme of
\cite{Salpeter:1967zz} is followed).
The thermal and electric contribution to the change in the bulk
modulus is nearly the same as one follows the pressure gradient.
Over ninety percent of a stars interior (all but the near surface)
can be described well by  a constant polytrope index that averages somewhere between
$4/3$ (thermal) and $5/3$ (electrical). 
As seen in this work the sensitivity of the polytrope index is
small enough that this $1/3$ difference has little effect on the final results.}

{The EOS developed is an unsophisticated example
of the power of this variable polytrope method.
At its foundation is the equivalency between the polytrope 
index and the important elastic variable,
the pressure derivative of the bulk modulus.
Other planet formation methods include a 
universal electric EOS in interplay with
thermal effects~\citep{Stacey2,Valencia2006545,Sotin,Grasset:2009xq,Leconte,
0004-637X-744-1-59,Wagner, Zeng2013} but they vary in their approach.
This technique has as its centerpiece one elastic variable from which all dynamic interactions
(electric, thermal, magnetic, nuclear) can contribute and thus one prescription to control input into
the differential equations.} 
This EOS has
the ability to mature
alongside its elastic variable polytrope index as we will discuss further in Sect.~\ref{S7}.

\section{Application to Planets}\label{S6}

We now apply our method to planets of our solar system. For
planets the changing polytrope index is a must for all but the smallest
planets. One looming approximation that needs to be addressed is the use
of cold isothermal material curves to describe planets with hot interiors.
We will address this in Sect.~\ref{S7}, but for now let us proceed assuming
that this simplification is feasible.

\begin{figure*}
\includegraphics[width=120mm]{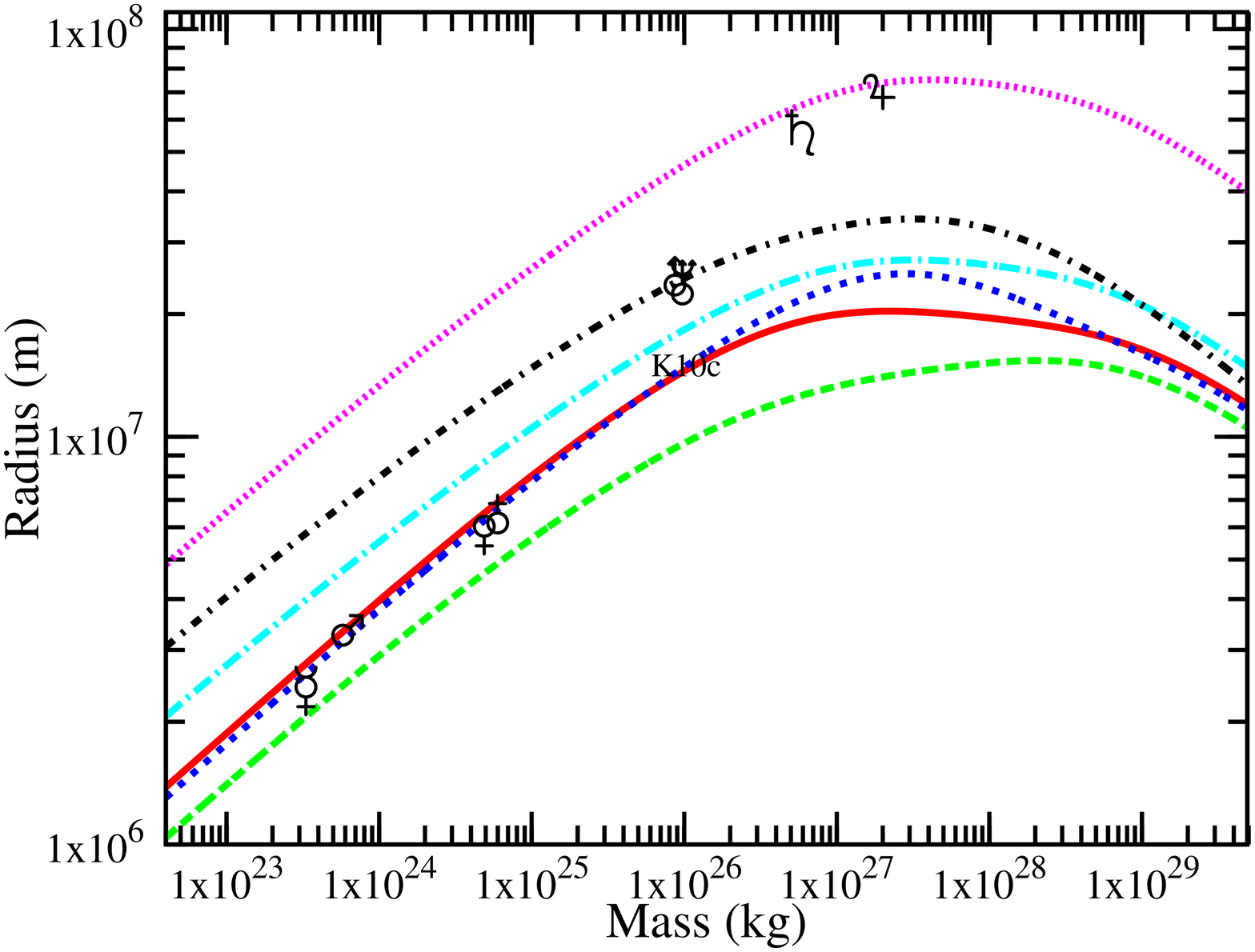}
\caption{(color online) The legend of the six lines are the
same as for Fig.~3, six materials that are common in planetary interiors. 
The symbols run vertically from top to bottom as
Jupiter, Saturn, Uranus and  Neptune (on top of each other),
Earth and Venus (on top of each other), Mars, and Mercury. 
{The symbol
k10c stands for exoplanet Kepler 10c which is a super-earth~\citep{0004-637X-789-2-154}.}}
\label{fig4}
\end{figure*}

In Fig. 4 we solve the modified Lane-Emden equation with a variable index
for the same six materials of the two previous figures except now we plot
mass versus radius.
This is a popular figure in planet modeling papers dating back to
at least~\cite{Zapolsky-1969}.
For comparison 
we designate the positions on the figure for the planets of the solar system.
Described from 
vertical top to bottom: Jupiter, Saturn, Uranus and Neptune (almost on top of each other), 
Earth and Venus (almost on top of each other), Mars and Mercury.
They are all placed close to their line of significant composition. Jupiter is close to 
hydrogen, Saturn has slightly more helium than Jupiter.
Uranus and Neptune are between water and helium,
Earth, Venus, Mars are between iron and the silicates. Mercury, the most metallic is closest
to iron. These logical results give confidence to the validity of the method. 
 {Examining the previous results of 
Zapolsky and Salpeter~(\citeyear{Zapolsky-1969}) and
Seager et. al~(\citeyear{Seager:2007ix}) we have very similar results where the radius hits a 
maximum for all materials at a mass somewhere between one thousandth and
one hundredth  of a solar mass.
This is not entirely surprising since all three studies 
used at extreme high densities the theory found in 
\cite{Salpeter:1967zz}. Our variable polytropic model does add additional physical insight into this
maximum. It is well known that the traditional polytrope has a fixed radius for the $n=2$ 
case~\citep{Chandrabook2}, the radius grows with mass for fixed $n>2$ and the radius shrinks 
with mass for fixed $n<2$. This maximum of Fig. 4 represents the variable polytrope analog to this 
fixed radius case. All maximum of Fig. 4 occur within a small range, $1.78< n(\rho) < 1.92$.
At the same time
this result shows little correlation with the vacuum density ($\rho_0$), bulk modulus ($B_0$), 
and pressure
derivative of the bulk modulus ($n_0$) for all six materials.
The importance of the $n\approx 2$ range is also evident by examining the change of sign in the
second term of
Eq.~(\ref{eq7})
which is our final version of the variable polytrope 
Lane-Emden equation. The general trends of Fig. 4 also validate 
the conclusions of \cite{0004-637X-665-2-1413} which put maximum radial limits on rocky planets thus
making it possible to separate the ice giants from the rocky giants. For an exoplanet super-earth
example we put Kepler 10c also on Fig. 4 (labeled 'k10c'). 
This planet is the mass of Neptune, extremely
hot at the core, and very dense~\citep{0004-637X-789-2-154}, our
temperature independent theory still places the planet in the correct bulk composition range.}

\begin{figure*}
\includegraphics[width=120mm]{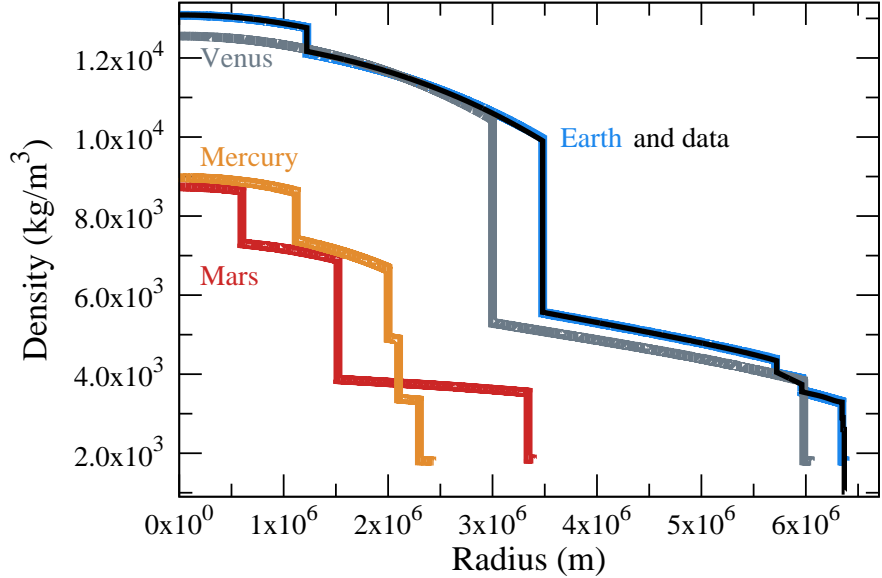}
\caption{(color online) Mercury, Venus, Earth, and
Mars density-radius profile predictions using our EOS. The
newer Reference
Earth Model~\citep{kustowski08} seismic data is plotted as a black line.
The details of construction are in the text.} \label{fig5}
\end{figure*}

Turning now to specific models for the four rocky planets in our solar system.
We constrain the boundary interfaces of these multi-layered planets
by substituting for the compressibility, $K$, in Eq.~(\ref{m3}) our variable index assumption
\begin{equation}
K = \frac{\rho_0^{n(\rho)} K_0 h(\rho)}{\rho^{n(\rho)}}. \nonumber
\end{equation}
{Using Eq.~(\ref{le4}),} we are then left with a constraint on the derivative at each boundary
\begin{equation}
\frac{d\rho}{dr} = -\frac{\rho_0^{n(\rho)}\;K_0 \;h(\rho)}
{\rho(r)^{n(\rho)-2}}\;g(r).\label{layer2}
\end{equation}
The method of construction is illustrative. We assume the
material makeup of the layers and the size of each layer, we then also assume the core
pressure. There are no further adjustable parameters. The materials dictate
the parameters in the equations of state, the core pressure determines the starting 
core density (by using Eq.~(\ref{ourpressure})), 
and the interface constraint defines the density values and their first derivatives at the
boundaries uniquely, the program
runs from the core ($r=0$) to the surface solving the differential 
equation of Eq.~(\ref{eq7}) and using the EOS as summarized in Eq.~(\ref{eos1}).
Numerically this differential equation is very stable and solvable with relative 
ease, there are no variables that gave us numerical difficulty.
Table~\ref{T3} contains our pertinent values for these multi-layer constructions. 
The materials at each layer are picked by consulting the best 'general knowledge'
about each of the layers and estimating their average density, bulk modulus, etc. 
Since the rocky planets are relatively small the significance of 
$A_{mean}$ and $Z_{mean}$  is minimal since their dependence stems from the
extremely high pressure theory (see subsection \ref{hp} for more detail).
The method is a standardized way to put a multi-layered planet together.
As supplemental material we include our code for the calculation of 
Earth ({\it Variable PolyPy}), written in {\it Python}.

In Fig. 5 we plot these hypothetical density-radius profiles for the four rocky planets.
In Table~\ref{T6}
we have displayed the results for mass, radius,
mean density, and gravity at the surface. For all three models the theoretical
results are within a percent of experimental values. This technique provides a 
systematic method to build mult-layered planets. One of the benefits of using
this method is that the EOS has its foundation in the elastic bulk modulus and its
pressure derivative. We believe this is an asset that can be utilized to develop
EOSs of greater sophistication, we study this question in the next section.

\begin{table*}
\centering
\begin{tabular}{ |l || c |c|c|c|c|c|c|| }
  \hline
& $\rho_0$&$B_0$&$n_0$&$A_{mean}$&$Z_{mean}$&thickness&$P_C$\\
& $kg/m^3$&$GPa$&     &    &  &$m$   &$GPa$\\
\hline
  Mercury &           &         &    &&&&   \\
\hline
  core 1 & 7700 & 180 & 5.0& 55&26& $1.12E6$ & $4.04E1$\\
  core 2 & 6410 & 132 & 4.9& 55&26& $8.80E5$ & $2.67E1$\\
  core 3 & 4800 & 220 & 4.8& 44&21& $1.00E5$ & $5.63E0$\\
  mantle & 3320 & 200 & 4.1&36&18&  $2.00E5$ & $3.63E0$\\
  crust &  1800 & 160 & 4.0&30&15& $1.36E5$  & $0.94E0$\\
\hline
  Venus &           &         &    &&&&   \\
\hline
  core  &  7475 & 160 & 4.95&47&22& $3.00E6$ & $2.95E2$\\
  mantle & 3800 & 190 & 4.0&36&18& $2.98E6$  & $1.22E2$\\
  crust &  1800 & 170 & 4.2&30&15& $7.11E4$  & $1.14E0$\\
\hline
  Earth &           &         &    &&&&   \\
\hline
  core 1 &   7550 & 171 & 5.0& 55&26& $1.22E6$ & $3.64E2$\\
  core 2 &   6830 & 140 & 5.0& 55&26& $2.26E6$ & $3.29E2$\\
  mantle 1 & 3950 & 190 & 4.4& 36&18& $2.23E6$ & $1.36E2$\\
  mantle 2 & 3650 & 205 & 4.1&36&18&  $2.51E5$ & $2.36E1$\\
  mantle 3 & 3270 & 135 & 3.8&36&18&  $3.84E5$ & $1.37E1$\\
  crust   &  1800 & 160 & 4.0&30&15&  $3.72E4$  & $0.66E0$\\
\hline
  Mars &           &         &    &&&&   \\
\hline
  core 1 & 7500 & 180 & 5.0& 55&26& $6.00E5$ & $4.04E1$\\
  core 2 & 6100 & 130 & 4.9& 47&22& $9.20E5$ & $3.66E1$\\
  mantle & 3530 & 200 & 4.2&36&18& $1.82E6$  & $2.20E1$\\
  crust &  1850 & 160 & 4.1&30&15& $5.10E4$  & $0.35E0$\\
  \hline
\end{tabular}
\caption{
Mars in two steps, Jupiter and Uranus in three steps.
The variables $\rho_0,B_0$, and $n_0$ would be the values
of those observables at zero vacuum. Thickness is the
radial size of the core or mantle part, $P_C$ is the
assumed pressure at the largest depth for each layer.}
\label{T3}
\end{table*}

\begin{table}
\centering
\begin{tabular}{ |l||c|c|c|c| }
  \hline
       &  mass    & radius   & mean density& surf. gravity\\
       & $kg$     & $m$      & $kg/m^3$    & $m/s^2$   \\
  \hline
  {\bf Mercury}&         &         &       & \\
  calculation &$3.29E23$ & $2.44E6$& $5.43E3$&3.70 \\
  experiment  &$3.30E23$ & $2.44E6$&$5.43E3$&3.70\\
  \hline
  {\bf Venus}&         &         &       & \\
  calculation &$4.83E24$ & $6.05E6$&$5.20E3$&8.80 \\
  experiment  &$4.87E24$ & $6.05E6$&$5.24E3$&8.87\\
  \hline
  {\bf Earth} &         &         &       &  \\
  calculation&$5.98E24$ & $6.38E6$&$5.49E3$&9.80\\
  experiment&$5.97E24$ & $6.37E6$&$5.51E3$&9.80\\
  \hline
  {\bf Mars} &           &         &      & \\
  calculation & $6.38E23$ & $3.39E6$ & $3.91E3$& 3.71\\
  experiment&   $6.42E23$ & $3.39E6$ & $3.93E3$& 3.71 \\
  \hline\
\end{tabular}
\caption{The rocky planets of our solar system
observable predictions using  our EOS. The experimental results
are taken from the {\it National Aeronautics and Space Administration} website.}
\label{T6}
\end{table}

\section{Using the pressure derivative of the bulk modulus as the Equation of State Foundation}
\label{S6b}
The role of the 
pressure derivative of the bulk modulus ($B^\prime \equiv \frac{dB}{dP}$),
which we have shown to be the polytrope index, $n$,
has grown in importance in material science
science~\citep{Birch,Keane,JGRB:JGRB6242,0953-8984-1-11-002,earth,0953-8984-17-39-007,0953-8984-18-46-015}.
As strain theory continued to progress this elastic observable assumed a more 
prominent role. It is, frankly, a wonderful observable. It is dimensionless, of order
one (for all materials and pressures), and because it is proportional to the second
derivative of pressure-density it has tremendous sensitivity. This work 
has developed a robust technique where the variable polytrope index is equivalent to this
elastic observable, $B^\prime$. 

\subsection{The importance of \texorpdfstring{$\frac{dB}{dP}$}{dBdP} in material research}
\cite{Stacey1}  developed an EOS that used 
$B^\prime \equiv \frac{dB}{dP} = n$ 
as the starting point. This work gives a thorough 
history of this technique and rightly recognizes 
\cite{Murnaghan15091944} and \cite{Keane} as instrumental
in the development of the Stacey EOS. Stacey and Davis
strongly believe that $B^\prime$ should be the center-piece of a
universal EOS~\citep{Stacey2,Stacey3,Lal,Singh-2012}. 

Calculating from first principles the electronic contribution to the bulk modulus
is well established in the literature~\citep{Wills}. The ability for modern 
material theorists to develop excellent $B^\prime$ theories from density functional
theory is not in doubt~(\cite{PhysRevB.82.144110} for example). 
There have also been some
recent high pressure models which include temperature as a direct extension of 
the pressure equation~\citep{Sotin,wagner:peer-00786873,Wagner}.
Our calculation technique would use alternative methods which show the thermodynamic
connection between $B^\prime$, the Gr{\"u}neisen parameter~\citep{Stacey2,Stacey3,Shanker} 
and temperature~\citep{Stacey2,Stacey4}, \cite{Sotin} gives a good
overview of both techniques.
The elastic constants,
such as $B^\prime$, are a natural environment 
to develop temperature dependence as shown in
a variety of research at low pressures~\citep{JGRB:JGRB15234},
first principle calculations of copper~\citep{PhysRevB.65.064302},
magnesium oxide~\citep{2005JGRB..110.5203L} or
high pressure iron~\citep{Xian10} and high pressure salt~\citep{Singh-2010}.
We will examine this potential by adding a primitive temperature dependence  in Sect.~\ref{S7}.
Phase changes would also have a natural place in a 
foundational $B^\prime$ EOS.  The discontinuities
created by phase changes would be more
significant in the bulk modulus and its derivative than in a pressure-density
curve.
The importance of magnetic effects has also recently been studied~\citep{PhysRevB.87.115130}
and similarly the $B^\prime$ equation has been shown to be a natural place to add this 
interaction~\citep{PhysRevB.82.144110}.
{Classical solid state theory
may also be of use. The intermediate polytrope
expression we derived in the context of the fixed
polytrope index, $\rho^nK = h(\rho)\rho_0^nK_0$ (Eq.~(\ref{m1})) in
Sect.~\ref{S4}, has some remarkable similarities to the functional form of the
repulsive coulomb potential informed by Born-Madelung theory~\citep{kittel}.}

\begin{figure*}
\includegraphics[width=120mm]{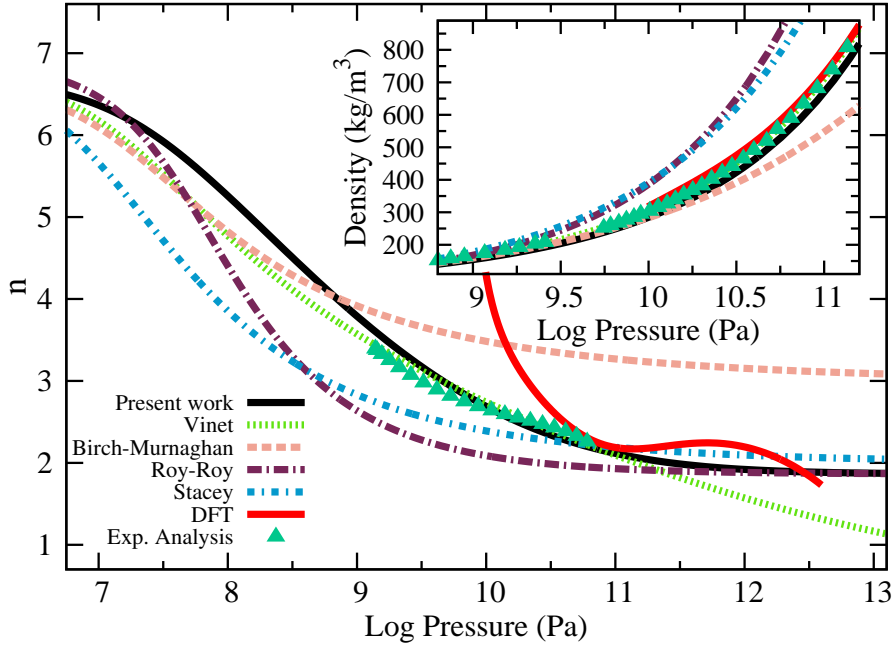}
\caption{(color online) A plot of $n\equiv\frac{dB}{dB}$ for cold-curve molecular
hydrogen using a variety of EOSs. The solid black 
line is the present calculation, the green dotted line is the EOS of 
Vinet~\protect\citep{JGRB:JGRB6242,0953-8984-1-11-002},
the pink dashed line is the third order EOS of Birch-Murnaghan~\protect\cite{Birch}. 
We also
present the EOS of Stacey~\protect\citeyear{Stacey1} (blue dash - dot 
fixing the infinite $\frac{dB}{dP}$ at 2)
and the EOS of Roy-Roy~\protect\citep{PSSB:PSSB125,0953-8984-11-50-328} is depicted
by a purple long dash - dot. A recent
density functional theory (DFT)  developed for molecular hydrogen~\protect\citep{Geng12} in red. 
All EOS (except the DFT)
have fixed $\rho_0=79.43 kg/m^3$, $B_0 = 0.162GPa$. The $n_0$ values are 
within 3\% of each other (ours, Birch-Murnaghan is $n=6.7$, Vinet is $n=6.813$
and we have set Roy-Roy and Stacey to $n=6.9$). Note the dramatic
differences in the function $n(P)$ and its effects on the density-pressure
relationship at extreme pressures.  The experimental analysis, in 
blue triangles, was done by 
taking 2nd derivatives of experiments done by \protect\cite{Loubeyre}} 
\label{fig6}
\end{figure*}

\subsection{Comparisons of \texorpdfstring{$\frac{dB}{dP}$}{dBdP}}
As stated in Sect.~\ref{S4} we found that
\begin{equation}
n(\rho) = \frac{dB}{dP}=B^\prime=A_0\left(\frac{\rho_0}{\rho}\right)^{A_1} + A_2 \nonumber
\end{equation}
works well. It is a respectable EOS for pressures up to
$10^{13}$ Pa and it can be extended to higher pressures by attaching it to quantum
mechanical models as also discussed in Sect.~\ref{S4}. 
This EOS was cultivated  by examining the functional form of 
$\frac{dB}{dP}$ for the Vinet
and Birch-Murnaghan EOS~\citep{Birch,earth} and also analyzing
pressure-density relationships found in 
previous work~\citep{0953-8984-17-39-007,0953-8984-18-46-015,Seager:2007ix,0004-637X-744-1-59}
including first principle calculations~\citep{French09,Hermann29122011,Geng12}
and acquired numerical second derivatives. This
was instrumental in developing the intermediate pressure ranges ($10^{11} - 10^{13} Pa$) which
older EOS often do not fit well. Our result is entirely empirical but it meets
the criteria that it is using $B^\prime$ as its foundation, that it is
analytically simple, and that it strives to go at least an order of magnitude
higher than the earlier EOSs.

The analysis included taking analytical second derivatives of EOSs using
\begin{eqnarray}
B &=& \rho\frac{dP}{d\rho} \nonumber\\
n &=& \frac{dB}{dP} = B^\prime=\frac{d}{dP}\left(\rho\frac{dP}{d\rho}\right) \nonumber \\
&&\mbox {so } \frac{dB}{B} = \frac{n}{\rho}d\rho. \label{axal}
\end{eqnarray}
The analytical results for the Vinet,
Birch-Murnaghan~\citep{Birch,earth}, \cite{Stacey1},  ~\cite{0953-8984-11-50-328}, 
and an EOS for
hydrogen derived from density functional theory~\citep{Geng12} are shown for
completeness in Appendix~\ref{CC}, these 
results are non-trivial fractions. One of the 
goals of this work 
was to develop a analytical formula for $\frac{dB}{dP}$
which is simpler than  the two
most used universal equations, the Vinet and Birch-Murnaghan,
yet at the same time mimic their behavior.

We portray $\frac{dB}{dP}$ in Fig.~\ref{fig6} for  molecular 
hydrogen as a function
of pressure for a variety of EOSs and in the inset
we show the pressure-density result. This figure is similar
to Figs. 2 and 3 but now the focus is on the behavior
of the polytrope index so the plot proper and inset have been switched. 
We choose hydrogen because it offers
challenges at relatively low pressures because of its high compressibility.
It was seen in Fig. 2 that a standard Vinet EOS does not do well as one 
approaches pressures of $10^{13} Pa$, and Fig.~\ref{fig6} gives the 
probable cause, the Vinet becomes soft as it heads towards a lower 
asymptote for $n$ than any of the other EOSs do.
In contrast, the 
complete failure of the Birch-Murnaghan EOS with these 
chosen parameters is because the asymptote for $n$ is higher than the rest
of the EOSs. All the EOSs (except the
DFT by~\cite{Geng12} which is valid in a limited range only) 
approach a constant at extreme pressures but they
disagree at what the infinite pressure constant ($B^\prime_\infty$) should be.
Analysis of the importance of this asymptote is examined in good detail in the
research of Stacey and Davis~\citep{Stacey1,Stacey2,Stacey3}.
The Birch-Murnaghan, in
Fig.~\ref{fig6}, has the highest $B^\prime_\infty$ of
$3$. The Vinet has the lowest, $B^\prime_\infty = 2/3$, 
the rest are  $5/3 \le B^\prime_\infty \le 2$. Our theory,
because it connects to the Thomas-Fermi-Dirac high pressure theory
has as its asymptote $B^\prime_\infty=5/3$, we set Stacey's EOS
to $B^\prime_\infty=2$ (they treat it as an adjustable parameter).

Note the peculiarity of the density function theory~\citep{Geng12}.
The authors choose a pressure EOS (see Appendix~\ref{CC}) which fit
only the pressures applicable to their theory ($1\times 10^{10}$ Pa
to $4\times 10^{12}$ Pa). When derivatives are taken to 
produce $\frac{dB}{dP}$ the results are complicated and quickly
unstable at the extremes (the ends of the plotted curve head towards infinity).
This insensitivity provides an additional lesson on how it
is easy to fit the density-pressure results and yet have faulty derivatives, an authoritative
EOS will fit the pressure-density, bulk modulus-density and bulk modulus pressure
derivative and density profiles. 

The experimental density-pressure profile comes from \cite{Loubeyre}. The 
technique we used to analyze this data to produce the polytrope index, as depicted in
Fig.~6, is detailed in  Appendix~\ref{CC}. To
summarize, we fit a density-pressure curve to a high order
polynomial and then take two derivatives, using Eqs.~\ref{axal},
to achieve a good analysis independently, directly from the data,
of $n\equiv\frac{dB}{dP}$. We urge that this technique be used by experimentalists
to further analyze their high precision 
data. With a good experimental set of density-pressure
data one can also determine the bulk modulus and the pressure derivative
of the bulk modulus over the same range. If one does want to 
extrapolate to vacuum pressure using a universal EOS, this information will help
produce an authoritative EOS and further constrain the parameters chosen. 
A similar technique and conclusion was proffered by
\cite{PhysRevB.68.064112}.

It has been argued by~\cite{Stacey2} that $B^\prime_\infty > 5/3$
but not necessarily $B^\prime_\infty = 5/3$. The argument is that
phase changes and proton number dependence make it difficult to rectify
that all materials will approach the same asymptote.
We also believe this is an open question but still found
it convenient to set it to the 
Thomas-Fermi-Dirac value of $5/3$~\citep{Salpeter:1967zz}.
Yet we recognize that the present work and others~\citep{Seager:2007ix}
have found that this asymptote is not reached quickly. With the additional
insight
that the polytrope index is the derivative with respect
to pressure of the bulk modulus we look at the interior of stars for 
guidance. All stars, at the core, have an index $<2$. All stars are
under extreme electronic, nuclear, and thermal pressure, thus indicating
that the contributions to $B^\prime$ from these sources are all under 2
in the extreme pressure limit.

Too often the universal EOSs are used
to put limits on the bulk modulus and its derivative at zero
pressure ($B_0,B^\prime_0$). All reasonable EOSs will fit any pressure-density
curve if the parameters are adjusted enough as 
discussed in ~\cite{Stacey2}, what Fig.~\ref{fig6}
shows is the disparity that exists between the EOSs on the functional
form and the asymptote of $\frac{dB}{dP}$. This observable needs to
be measured or analyzed experimentally  
as well as theoretical approaches from first
principles~\citep{PhysRevB.68.064112}. The observable should
be ascertained
independent of any universal EOS
so the best
functional forms can be developed. 
Molecular hydrogen, helium, and water, because of their higher compressibilities,
are a good place to start. 
{These informed results would catalyze the development
of EOSs. With a trusted
functional form, which mimics experiment for pressure-density,
bulk modulus-density, and especially $B^\prime$-density (which is
sensitive even at low pressures), there will
be more faith in the physical
interpretation of the parameters being adjusted.
To constrain the universal
functional form beyond the pressure-density profile  would have significant impact  on
calculations of larger planets. In the next section we use these objectives on fitting 
Earth using our EOS.}

\section{Analysis of planet Earth with our EOS}
\label{S7}
The Earth is our best pseudo-static 
laboratory for studying materials under high
pressures. In this section we analyze the Earth using different
constraints on our EOS to better elucidate its strengths and weaknesses.

By using the seismic data of planet Earth we now constrain, with our EOS,
the material of the inner core, outer core and mantle using 
the pressure-density profile, and the density-radius profile. 
The best results still achieve an excellent fit and
reveal a linear analysis similar to the inset of Fig. 1,
($\rho_0^{n_0}K_0$ vs. $n_0$) An example calculation was shown in
Fig.~\ref{fig5} and the code which produced this calculation is included
as supplementary material.
The linear equations
for the inner core, outer core, and lower mantle of the Earth 
are as follows: \\
  inner core: $\log_{10}(\rho_0^{n_0}K_0)= 3.994n_0-11.813$\\
  outer core: $\log_{10}(\rho_0^{n_0}K_0)= 3.977n_0-11.836$\\
  lower mantle: $\log_{10}(\rho_0^{n_0}K_0)=3.702n_0 - 11.717$ \\
A reminder that the linear analysis constrains $\rho_0^{n_0}K_0$ but does not guarantee
that every point on the line will be a good fit, but all good fits are 
very near that line. We offer no explanation for this relationship
but that its success does speak to the validity of universal EOSs and that the 
three variables ($\rho_0,K_0,n_0$) are not independent. An 
oversimplification in our technique is that the experimental material
fits were all to cold curves, we will address this assumption in the next
subsection.

\subsection{Temperature Analysis}\label{S7b}
\cite{Seager:2007ix} have argued
that temperature dependencies are significant but manageable (less than 10\%
in a pressure-density profile). We agree and the reasonable physical
observables predicted in Fig.~\ref{fig4} and Fig.~\ref{fig5} 
for the rocky planets
exemplify the power of using approximate cold curves.
By using the seismic data of planet Earth we may now, with our equation
of state, constrain further 
the material of the inner core, outer core and mantle and examine the role
of temperature dependence for our EOS.

As one moves from the surface of a planet to its core the polytropic
index, $\frac{dB}{dP}$, moves from close to its vacuum value to a lower value. The
effect of temperature is to slow this descent as one immerses into the hotter interior
in an approximately adiabatic path.
For all of our calculations above, which were cold curves, we fixed the
product and sum of our parameters to $A_0+A_2 = n_0$ and $A_0A_1=n_0$. The sum
is required for all curves, the product was a fortuitous choice 
suggested by Roy-Roy~(\citeyear{0953-8984-18-46-015}) which worked
well for cold curves, it controls the relative size of the second
derivative
\begin{equation}
\frac{dn}{d\rho} =-A_0A_1\frac{\rho_0^{A_1}}{\rho^{A_1+1}}
\end{equation}
(see appendix \ref{oureos} for more detail). The cold curves $A_2$ was
set by a numerical secant search that connected our cold
curve EOS to the Thomas-Fermi-Dirac EOS, the values for $A_2$ was always between 
1.75 and 2.1.
So by increasing the value of 
$A_2$ we decrease the magnitude of this derivative thus providing a simple method
to include temperature. The procedure
is simple: assume a reasonable material with a given vacuum density ($\rho_0$), 
bulk modulus ($B_0$), and first pressure 
derivative of the bulk modulus($n_0$). The 
constraints on $A_0, A_1,$ and $A_2$ are that $n_0 = A_0+A_2$ and $n_0=A_0A_1$. We treat as free parameters
$\rho_0,B_0,n_0$ and $A_2$ and adjust them using a Powell method 
(constraining $A_0$ and $A_1$ by $A_0+A_2 = A_0A_1 =n_0$). 
So we are 
assuming that at the surface of the Earth (vacuum pressure) 
the density, bulk modulus, first derivative with respect
to pressure, and all higher order derivatives have the same value
irregardless of whether the isothermal or adiabatic path is chosen. But as one moves
into the interior $B_{adiab} > B_{isotherm}$ thus the elastic derivatives are also disparate.
In \cite{Stacey2,Stacey4} one finds a thermodynamic derivation 
that provide a mapping 
\begin{equation}
B_{adiab} = B_{isothermal}(1 + \gamma\alpha\Delta T),\label{bulky}
\end{equation}
where $\gamma$ is the Gr{\"u}neisen parameter and $\alpha$ 
is the volume expansion coefficient
{(which are related to the bulk modulus and
its pressure derivative)}. The reference temperature for $\Delta T$ is low, the 
temperature at the surface of the Earth. This ratio of
$B_{adiab}$ over $B_{isothermal}$ remains close to normal, in 
\cite{Stacey2} it never goes above 1.10 for planet Earth. This is a substantial
piece of evidence that the
isothermal cold curves are a good approximation for the interior of the Earth. 

As
$B_{adiab}$ increases faster than $B_{isothermal}$ as the pressure increases, this
implies directly that 
$n_{adiab} \equiv \frac{dB_{adiab}}{dP} > \frac{dB_{isotherm}}{dP}\equiv n_{isotherm}$
Pragmatically this is not too difficult, we must have an $A_2$ 
greater than the cold isothermal curves ($>2.2$)
and so we vary $\rho_0,B_0,n_0$ and $A_2$ to fit the interior of the Earth. This fitting has 
only one more adjustable parameter,$A_2$, than the cold isothermal curves. We will
label this primitive 
temperature dependent method 'model VII', after the section from which it was introduced.

\begin{figure*}
\includegraphics[width=120mm]{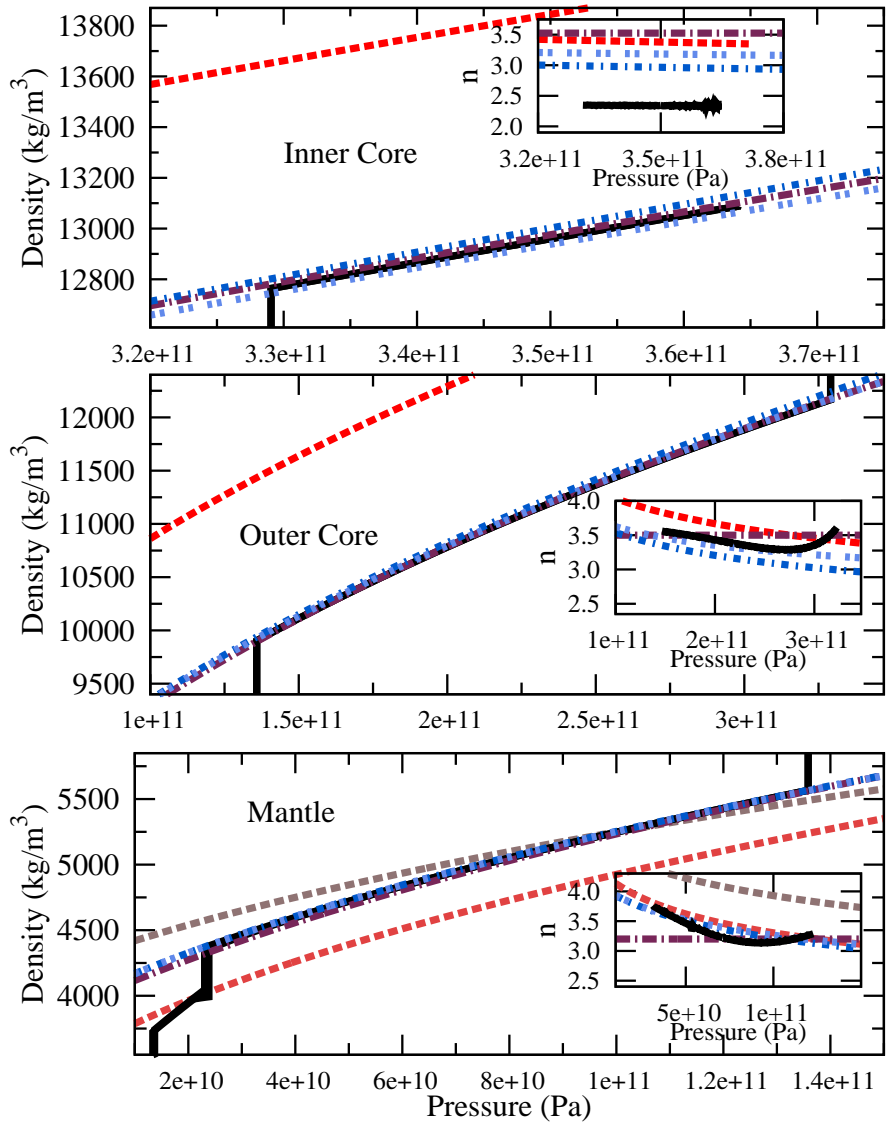}
\label{fig7}
\caption{(color online) This figure is in three panels: the top
referencing the inner core, the middle the outer core, and the bottom
is the inner mantle. In black,
significantly buried under the theoretical calculations, is the
newer Reference
Earth Model~\citep{kustowski08} (REF). The colored lines are 
all calculations of ours with different sophistication, all
parameters are given in Table~\ref{T1b}. The three models are named
from the section from which they were introduced. 
Model (VII -- two light blue 
short dashed) includes
an adjustable second derivative ($\frac{d^2B}{dP^2}$) to simulate
the effect of temperature in the interior of the Earth. 
Model (IV - short dark blue dashed-dot) is the
same material as (1) but the second derivative functional form has now been fixed to
the cold curve value effectively removing the temperature dependence. 
Model (III -- purple long dash-dot) is the least sophisticated fixed polytrope
index (($\frac{d^2B}{dP^2} = 0$). For reference the cold curve for hcp iron is
included for the cores (dashed red) and magnesium oxide (dashed pink) 
and silicon dioxide (dashed brown) is
included for the mantle.
The inset contains a depiction of the function $n(P) \equiv \frac{dB}{dP}(P)$ for our three
models along with experimental analysis direct from the seismic wave data of
the
newer Reference
Earth Model~\citep{kustowski08} (REF) depicted by the thick black line.
}
\end{figure*}

In Fig.~7 we examine the inner core, outer core and mantle of the Earth in
more detail than Fig.~1 and Fig.~\ref{fig5}. We begin by plotting the seismic results, as we did
in Fig.~1, except now we plot the pressure-density profile (sold
black line).
Since the data sets are large and precise they are informative.
Not only do they contain
the pressure-density but they also implicitly contain 
the derivatives. Here we calculate the bulk modulus
directly from the $s$ and $p$ wave velocities (see \cite{Stacey2} for
example) and we then calculate the pressure 
derivative of the bulk modulus 
numerically which is represented by the black line on the inset. 
We also plot three models from this work (named after the section 
number from which they were introduced):
a good fit using the fixed polytrope index from Sect.~\ref{S3} 
(model III,$\frac{d^2B}{dP^2}\equiv 0$,
purple horizontal line), 
a good fit using a variable polytrope index 
from this section (temperature dependent,
model VII, light blue dashed line), 
and the same material as model VII with the temperature
dependence removed, thus a isothermal cold curve  as described in Sect.~\ref{S4}
(model IV, dark blue dashed line).
For reference we also plot,
from Sect.~\ref{S4},  the cold curves for
iron, silicon dioxide, and magnesium oxide at those pressures. 
Model VII was fit simultaneously to the density-pressure curve (main graph), 
the bulk modulus, the density-radius curve (Fig.~1) 
and the pressure
derivative of the bulk modulus-pressure curve (inset).
This last constraint, depicted in the inset, shows the wonderful dynamics that
are exhibited within the interior of the Earth. The outer core
and mantle do not adhere to the  
monotonically decreasing pressure derivative of the bulk modulus
of our cold curve or primitive temperature models but as the 
layer boundary approaches the second pressure derivative of the bulk modulus
changes sign (the material's rate of hardness growth increases!) as
surmised by the analysis of the seismic data. The mantle and outer core have shown to be 
wonderful examples of phase-changes, non-adiabatic regions, and convection
\citep{alboussiere2010,Katsura2010212,Matyska20111} as the seismic waves slow
as they near these boundary layers. The
simple form of our EOS cannot describe this dynamic and so we 
hope only to stay close. Luckily the sensitivity to 
the pressure derivative is manageable. 
{This polytrope method forces the user to contemplate beyond the
density-pressure profiles to constraining the derivatives thus making
it an ideal tool for bulk composition studies in the interior of a rich dynamic planet.} 

Our simple temperature dependence does an adequate job describing the 
changing polytrope index for the mantle and the outer core, 
especially away from the inner boundary layer. The direction
is correct, the magnitude is of the right order for the thermal pressure
and the bulk modulus differential of Eq.~(\ref{bulky}) 
($B_{adiab} / B_{isothemal} \approx 1.01$ for the mantle and $\approx 1.04$ for
the outer core). Most importantly
we can describe the functional form of the polytrope index adequately.
The inner core has the right direction and magnitude also
($B_{adiab} / B_{isothemal} \approx 1.04$ for the inner core) but we fail to
find an adequate function to match the low experimental analysis of the 
interior polytrope index($n \approx 2.3$!).
The most obvious potential source of error is that the functional 
form of our temperature dependence 
does not have the complexity to adequately describe the adiabatic curve relative
to the cold temperature curve at inner core pressures. Recent experimental
research~\citep{Tateno15102010,Anzellini26042013} also indicates the presence of a 
phase change near the geothermal which a universal EOS cannot describe.
We also could easily be missing important magnetic~\citep{PhysRevB.87.115130} or rotational effects.
 {Again this method naturally extends the variables open to sensitivity
examination thus providing a systematic tool to examine the successes and failures of any chosen EOS.}
We do have some faith in our estimation for $n_0$ of the core because
we do match the bulk modulus number at the core to a good precision.

\begin{table}
\centering
\begin{tabular}{ |l || c |c|c|c|c|c| }
  \hline
Model &$\rho_0$&$B_0$&$n_0$& $A_0$ & $A_1$  & $A_2$\\
from Sect.&   $[kg/m^3]$ & $GPa$ &&&& \\
\hline
Earth Inner Core & \multicolumn{6}{c|}{}\\
\hline
Sect. VII & 7563.0&196.39& 4.486 &1.736 &2.584 & 2.750\\
Sect. IV & 7563.0&196.39& 4.486 &2.369 &1.893 & 2.117\\
Sect. III&7372.6& 194.74&3.525&0.0 & - & 3.525\\
Fe &8300.0&165.0&5.150&3.080& 1.672 &2.070\\
  \hline
Earth Outer Core & \multicolumn{6}{c|}{}\\
\hline
Sect. VII& 6654.2&131.32& 4.748 &2.096 &2.266 & 2.652\\
Sect. IV & 6654.2&131.32& 4.748 &2.652 &1.791 & 2.096\\
Sect. III&6596.7& 153.33&3.500&0.0 & - & 3.500\\
\hline
Earth Mantle & \multicolumn{6}{c|}{}\\
\hline
Sect. VII& 3988.4&207.30& 4.104 &1.804 &2.275 & 2.300\\
Sect. IV & 3988.4&207.30& 4.104 &2.199 &1.867 & 1.905\\
Sect. III&3942.9& 216.85&3.200&0.0 & - & 3.200\\
MgO&3580.0& 157.0&4.371&2.465 & 1.772 & 1.904\\
SiO$_2$&4287.0&305.0&4.750&2.983 & 1.592 & 1.767\\
\hline
\end{tabular}
\caption{$\rho_0, B_0, n_0$ are chosen to be
a  good fit to the models of the Earth based on experiment.
Model (VII - two dashed light blue on Fig.~7)
is the most realistic allowing the parameters
$A_0$ and $A_1$ to vary to simulate temperature dependence.  Model (IV - dark blue
short dash-dot on Fig.~7) is the same material with the temperature dependence removed
and thus not as good fit.
The simplest
model (III - purple long dash-dot on Fig.~7) is a fixed polytrope index. The
fixed polytrope does not require the sophistication of the 
earlier models but it can still be reduced to a specific case
of the variable model by setting $A_0=0, A_2 = n$ and noting that
it is independent of $A_1$. For reference
 we depict the cold curve prediction for
$\epsilon$-iron in the core panels, we add magnesium oxide and
silicon dioxide for the mantle.}
\label{T1b}
\end{table}

The results of the three models are systematically logical. 
For the best fit with realistic
values for the vacuum density, bulk modulus, and pressure
derivative of the bulk modulus the adiabatic curve version
with an unsophisticated temperature dependence, model VII,  had good success as the listings show
in Table~\ref{T1b} and depicted in Fig.~7. This result implies
that this technique has the potential to
add temperature and any other dynamic
variables if the effect of these variables
on $\frac{dB}{dP}$ are known (profiles). Even with our simple adiabatic temperature 
model we were
able to fit the density-radius profile to $<$ 2\%, the density-pressure profile
to $<$ 1\%, the bulk modulus of the core to $<$ 3\%, and the pressure 
derivative of the bulk modulus for the outer core and mantle to $<$ 10\%,
the inner core we fit to 30\%. 

Recognizing that choosing
our EOS as authoritative (the best functional forms for the cold curves) is a 
source of possible systematic error (when our EOS errs it is likely to error on 
the hard side as depicted by Fig.~\ref{fig6})  as is our simple model for adiabatic
temperature dependence (the slope of the polytrope index-pressure curve 
is low for mantle and outer core), we choose to 
constrain the interior of the Earth
materials with conservative, fair errors: \\
{\bf inner core:} $\rho_0 = 7600 \pm 300 kg/m^3$, 
$B_0 = 190 \pm 40 GPa$, and $B_0^\prime = 4.7 \pm 0.5$\\
{\bf outer core:} $\rho_0 = 6800 \pm 300 kg/m^3$,
$B_0 = 130 \pm 30 GPa$, and $B_0^\prime = 4.8 \pm 0.3$\\
{\bf mantle:} $\rho_0 = 4100 \pm 200 kg/m^3$,
$B_0 = 200 \pm 30 GPa$, and $B_0^\prime = 4.2 \pm 0.3\;$ .\\
The constant $\rho_0^{n_0}K_0$, which plays a pivotal
role in this research, explains why the bulk modulus is
difficult to constrain. This constant is
extremely sensitive to the compressibility and its inverse, 
the bulk modulus. A small change in $\rho$ or $n$ will cause
a much larger change in the bulk modulus if this term is to stay near constant.

If the effects of temperature and other dynamic variables 
are not specified explicitly
then a reasonable material cold curve can be used instead and 
the error
will be very manageable, less than 5\% for a rocky planet like Earth as
detailed in the analysis of the ratio of Eq.~(\ref{bulky}) which we 
always found to be under 1.05 using our primitive temperature dependence model.
This trend is further bolstered by examining the difference between the density-pressure
curves of Fig.~7.
So we recommend the use of cold curves, with reasonable material observables,
as a very good approximation for the actual geothermal path that describes
the rocky planet under study.
However we do not recommend the reverse analysis of trying
to fit only the pressure-density
experimental data
with a cold curve to elucidate
the the vacuum elastic variables the results
will also be suspect and highly
dependent on the EOS used. 

By numerically analyzing the experimental seismic data we have constrained
our EOS further than most (fitting to pressure-density, density-radius, bulk modulus-pressure and
$\frac{dB}{dP}$ vs. pressure) and can thus eliminate many previous parameter
choices. We have also added a primitive temperature dependence
which has the correct direction and the right order of
magnitude. So we felt confident, with our model VII, to constrain the materials in the 
interior of the Earth with moderation. For  better analysis we would need a reasonable temperature,
magnetic, phase-change and convection  profile, prepared from first principles, which
would more tightly constrain the parameters. 

\subsection{Summary of Technique}
The method presented here is systematic, builds on history, and is able to
grow in complexity as the EOS becomes more complex.
To summarize this theoretical apparatus to solve the radius-density profile for a multi-layered planet:
\begin{itemize}
  \item  Develop a functional form or differential equation(s) 
for the polytrope index as a function of density. The dynamic variables that describe
this polytrope index function can be sophisticated and include a full 
thermal, electromagnetic, and nuclear profile:
$n(\rho) \equiv \frac{dB(\rho)}{dP}$ for an appropriate core material with a given 
$\rho_0,B_0,B_0^\prime \equiv n_0$ and core pressure.
  \item Solve analytically or numerically
for the weighting function:$\ln h(\rho) = 
\int_{\rho_0}^{\rho}\frac{dn}{d\rho}\ln(\frac{\rho}{rho_0})\; d\rho$
  \item Starting at the inner core ($r=0$) find the core density using the weighting function $h(\rho)$
and core pressure relationship, also
set $\frac{d\rho}{dr}= 0$. 
   \item Solve the Lane-Emden differential equation (Eq.~(\ref{eq7}))
to a chosen radius (or if the
surface, when $\rho(r) = \rho_0$).
   \item If another layer is desired choose an appropriate material with a given
$\rho_0,B_0,B_0^\prime$ then calculating  $n(\rho)$ and $h(\rho)$ as shown above.
   \item Choose an initial density for the next layer and a starting radius ($r= r_{final}$
of the previous layer).
   \item Find the initial derivative, $\frac{d\rho}{dr}$ by using Eq.~(\ref{layer2}) and
Eq.~(\ref{eq7}).
   \item Solve the differential equation to a chosen radius (or if the
surface, when $\rho(r) = \rho_0$). Repeat the last 4 steps if additional layers are desired.
\end{itemize}

If one uses a series of differential equations and profiles 
which produce numerical functions only for 
the polytrope index, $n(\rho)$, it would be preferred
to start at the surface of the planet and work 
inward since the calculus relations will also not be analytical. 
This would involve some more computational challenges, but they are not insurmountable as
\cite{Zeng:2008cz} have demonstrated. 

\section{Summary and Conclusion}\label{S8}

{In summary this work derived a variable polytrope approach. To investigate
the method we also developed a new universal EOS which is 
distinctive because it is a function of the polytrope index.}
This work first exhumed, in Sect.~\ref{S2}, that the historic polytrope index 
is
for all cases and materials equivalent to the derivative with respect to pressure of
the bulk modulus ($n=\frac{dB}{dP}$). This result is the foundation of the article, the 
further development that follows is inspired and motivated by
recognizing the physical implications of the polytropic index as an elastic constant thus
giving it a life beyond merely index status.
We solve the Lane-Emden differential equation for
plausible solutions to the density profile of the five layer interior
of the Earth. We find that there is a wide range of 
constant polytropic indexes ($n = \frac{dB}{dP}$) that agree with the seismic data of
the interior of the Earth.

In Sect.~\ref{S4} we expand our technique by making the polytropic index a variable.
Since the index has been shown to be equivalent to the derivative with 
respect to pressure of the bulk modulus ($n = \frac{dB}{dP}$) this is a natural
progression since all materials have a decreasing first derivative 
along the isotherm when the pressures become
very large. We re-derive the Lane-Emden differential equation for a variable polytropic index
in a consistent manner. We then introduce an empirically developed functional form for
our variable polytropic index, $n(\rho) = A_0(\frac{\rho_0}{\rho})^{A_1}+A_2$, which is our
submission for a isothermal universal equation of state. With this creation we can create analytical
formulas for pressure and bulk modulus as a function of density. We, more importantly, use our
variable polytropic index as input
in the Lane-Emden differential equation, we are able to make pressure-density profiles
for six materials which are common in the interior of planets which 
compare favorably to experiment. By connecting our results
to high pressure theory~\citep{Salpeter:1967zz} we are able to make theoretical calculations
up to $10^{18}Pa$.
We turn to the planets of our solar system in Sect.~\ref{S6} where we show that their
masses and radii compare favorably with mass-radius curves for the six common materials.
We then show that this technique can be applied to a multi-layered body as we fit our
technique to the rocky planets.

In Sect.~\ref{S6b} and Sect.~\ref{S7} we discuss the importance of
developing a universal EOS from $\frac{dB}{dP}$, we then compare our
functional form for $\frac{dB}{dP}$ to previous work and to experimental analysis. 
The importance of constraining a EOS fit beyond the mundane pressure-density
profile is demonstrated.
We show that our EOS has the
advantage of being simple and thus easy to grow in sophistication. 
One addition that is needed is the 
adding of temperature dependence, we do this in primitive form
and show a variety of results using the interior of the Earth as our laboratory.
We also verify earlier work that shows that temperature modification is a relatively
small effect, under 5\% or the Earth's density-pressure profile. Finally using
a variety of experimental analysis, we are able
to narrow the constituent vacuum densities of the Earth to 5\%, the bulk moduli to 20\%, and the
pressure derivative of the bulk moduli to 10\%.

Having revealed the polytrope index as the pressure derivative of the bulk modulus is
fortuitous for future research advancement. There are many first principle EOS theories
(density functional, quantum Monte-Carlo) which calculate the pressure derivative
of the bulk modulus~\citep{Wills,Xian10,Xun10,2005JGRB..110.5203L}.
The prominence of the
bulk modulus in this procedure may also lends itself to solid state Born-Madelung
calculations~\citep{kittel}.
By using the modified polytrope assumption of this work one also
has a straight forward procedure to solve the 
Lane-Emden equation for planets and stars using
more sophisticated equations of state. These methods may include
dynamic convective interactions, heat flow,
rotational dynamics, magnetic kinetics, and nuclear reactions 
 which would all contribute
to the overall pressure derivative of the bulk modulus, the variable polytrope index,
complementing a new generation of realistic
planetary and solar models.
{We hope that this technique 
strengthens the relevance of the polytrope method for planet and solar
research and continues what Lane, Eddington, and Chandrasekhar began
a century ago.}

\section*{acknowledgements}
{S.~P. Weppner and J.~P.~McKelvey would like to thank Don Vroon and Ray Hassard
for bringing them together and Rohan Arthur for comments. K.~D.~Thielen and A.~K.~Zielinski
would like to thank Eckerd College for an internal grant which funded them during
the summer of 2013. It is with great sadness that we report the passing of
J.~P.~McKelvey in July of 2014.}

%\def\href#1#2{#2}
%\def\urlinner#1{#1\endgroup}
%\def\url{\begingroup\def\do##1{\catcode`##1 12 }%
%  \do\\\do\$\do\&\do\#\do\^\do\_\do\%\do\~ \ttfamily \urlinner}
%\bibliographystyle{mnras}
%\bibliography{polytrope}

\appendix

\section{Relations between density and Compressibility from the Murnaghan EOS}
It was Stacey and Davis~\citeyear{Stacey2,Stacey3} who we have seen state
that in the context of the Murnaghan EOS the relationship 
\begin{equation}
\frac{B}{B_0} = (\frac{\rho}{\rho_0})^{B_0^\prime}, \label{a1}
\end{equation}
which is equivalent to our own intermediate results for the fixed
polytrope index (Eqs.(\ref{inter},\ref{inter2})).
The derivation of Eq.~(\ref{a1}) is straightforward but we include it
in some detail because the relation is used extensively in this work.

Assuming only a linear change in the bulk modulus
\begin{equation}
B = B_0+ B_0^\prime P, \label{a2}
\end{equation}
where $B_0^\prime$ is the fixed derivative with respect to pressure. One can then
insert the definition of the bulk modulus
\begin{equation}
\rho\frac{dP}{d\rho} = B_0+ B_0^\prime P,
\end{equation}
and integrating and describing the zero pressure boundary conditions 
by using a naught subscript, one gets
\begin{equation}
\frac{\ln(B = B_0+ B_0^\prime P)}{B_0^\prime} = 
\ln\frac{\rho}{\rho_0}+\frac{\ln B_0}{B_0^\prime}
\end{equation}
which reduces to
\begin{equation}
(1+\frac{B_0^\prime P}{B_0}) = (\frac{\rho}{\rho_0})^{B_0^\prime}.\label{a4}
\end{equation}
Finally using Eq.(\ref{a2}) to substitute $B_0^\prime P = B - B_0$ 
we have the desired relation, Eq.~(\ref{a1}).

The Murnaghan EOS is found by solving Eq.~(\ref{a4}) for pressure
\begin{equation}
P= \frac{B_0}{B_0^\prime}\left ( (\frac{\rho}{\rho_0})^{B_0^\prime} - 1 \right ). 
\label{murn2}
\end{equation}
The similarity of this equation to the polytrope is undeniable
\section{Our Empirical Equation of State}
\label{oureos}
We developed an equation of state from analysis of earlier empirical
equations of state. The major difference is ours uses the polytrope
index, the pressure derivative of the bulk modulus, as its 
place of development. It was first stated in the text proper as
Eq.~(\ref{eos1}), again
\begin{equation}
n(\rho) = \frac{dB}{dP}=A_0\left(\frac{\rho_0}{\rho}\right)^{A_1} + A_2. \nonumber
\end{equation}
The constants are chosen as follows:
$A_2 \rightarrow B^\prime_\infty$ (approximately 1.95 {\bf or}
connecting it to the results of \cite{Salpeter:1967zz}), 
$A_0 = B_0^\prime - A_2$, and
$A_1 = \frac{-B_0B_0^{\prime\prime}}{B_0^\prime - A_2}$  
where $ B^\prime_\infty$ is the asymptotic value at very
high densities, see Appendix~C for more details
In this work we 
set $B_0$, and $B_0^\prime$ to experimental
results and assume a universal ratio for $B_0^{\prime\prime}$
by setting
\begin{equation} 
A_1 = \frac{B_0^\prime}{B_0^\prime-A_2}= \frac{B_0^\prime}{A_0}, \label{roy}
\end{equation}
which we borrowed from Roy-Roy~(\citeyear{0953-8984-18-46-015}).
Now using these
fundamental relations
\begin{eqnarray}
B &=& \rho\frac{dP}{d\rho} \\
n &=& \frac{dB}{dP} = \frac{d}{dP}\left(\rho\frac{dP}{d\rho}\right) \\
&&\mbox {so } \frac{dB}{B} = \frac{n}{\rho}d\rho \\
\ln h &=& n\ln\frac{\rho}{\rho_0}+\ln\frac{B_0}{B}
\end{eqnarray}
in conjunction with Eq.~(\ref{m2})
\begin{equation}
\frac{dn(\rho)}{dP}\ln\frac{\rho}{\rho_0} =  \frac{d}{dP} \ln h(\rho). \nonumber 
\end{equation}
It can thus be shown
\begin{equation}
\frac{dh}{h}=\ln\frac{\rho}{\rho_0}\frac{dn}{d\rho}\:d\rho .
\end{equation}
Specifically applying these relations to our 
EOS as expressed in Eq.~(\ref{eos1}) we find
\begin{eqnarray}
n &=& \frac{dB}{dP}=A_0\left(\frac{\rho_0}{\rho}\right)^{A_1} + A_2 \\
B &=& B_0e^{M(\rho)}\left(\frac{\rho}{\rho_0}\right)^{A_2} \\
P&=& \frac{B_0e^{\frac{A_0}{A_1}}}{A_1}
\left [\left(\frac{\rho}{\rho_0}\right)^{A_2}E_n\left(\frac{A_1+A_2}{A_1}
,\frac{A_0}{A_1}\left(\frac{\rho_0}{\rho}\right)^{A_1}\right)  
- E_n\left(\frac{A_1+A_2}{A_1},\frac{A_0}{A_1}\right)\right ] \\
\frac{dn}{d P} &=&\frac{d^2B}{dP^2}
= -A_1\frac{(\frac{dB}{dP}-A_2)}{B}  \\
\ln h &=& \frac{A_0}{A_1}\left(\frac{\rho_0}{\rho}\right)^{A_1}\left(1+A_1\ln\frac{\rho}{\rho_0}\right ) 
- \frac{A_0}{A_1} \label{eos5} \\
h &=& e^{-M(\rho)}
\left(\frac{\rho}{\rho_0}\right)^{A_0{(\frac{\rho_0}{\rho})^{A_1}}},
\end{eqnarray}
noting that only $n(\rho)$ and $h(\rho)$ are needed to solve the Lane-Emden
differential equation, Eq.~(\ref{eq7}).  The functions in the exponents are given by
\begin{equation}
M(\rho) = \frac{A_0}{A_1}(1-(\frac{\rho_0}{\rho})^{A_1}) \nonumber \\
\end{equation}
The only equation which is non-analytical is pressure 
which contains a special integral function, 
the generalized exponential integral ($E_n$) which is defined as
\begin{equation}
E_n(n,x) = \int_1^\infty \frac{e^{-xt}}{t^n}dt.
\end{equation}

By using the assumption of Ref.~\citep{0953-8984-18-46-015},
we are equating the second derivative as
\begin{equation}
B_0^{\prime\prime} = -\frac{B_0^\prime}{B_0}.\label{eos2}
\end{equation}
This is true at the surface but we can find the general relation also for our EOS. 
First taking 
the derivative of the polytrope index with respect to density,
\begin{equation}
\frac{dn}{d\rho} =-A_0A_1\frac{\rho_0^{A_1}}{\rho^{A_1+1}} 
\end{equation}
then switching variables to a pressure derivative we have
\begin{equation}
\frac{dn}{dP}=-A_0A_1\frac{\rho_0^{A_1}}{\rho^{A_1+1}}\frac{d\rho}{dP} =
-A_0A_1\frac{\rho_0^{A_1}}{B \rho^{A_1}} = -A_1\frac{(\frac{dB}{dP}-A_2)}{B}.\label{eos3}
\end{equation}
By doing a substitution for $A_1$ in terms of $A_2$ (using Eq.~(\ref{roy}))
in Eq.~(\ref{eos3}) we have finally
\begin{equation}
\frac{dn}{dP} = \frac{d^2B}{dP^2} = -(\frac{B_0^\prime}{B_0^\prime-A_2})
\frac{(\frac{dB}{dP}-A_2)}{B},
\end{equation}
which has an interesting symmetry. 
This relationship for the second derivative has been studied
in some detail recently in~\cite{Singh-2012}.

\section{Parameters chosen for materials}
\label{param}
In the text proper we used six materials commonly found in the interior of planets
in our solar system.
For completeness and reproducibility we include our parameters for the 
lower
pressures ($<10^{13}$ Pa) in Table~\ref{T1}.

\begin{table}
\centering
\begin{tabular}{ |l || c |c|c|c|c|c| }
  \hline
material &$\rho_0$&$B_0$&$n_0$& $A_0$ & $A_1$  & $A_2$\\
&   $[kg/m^3]$ & $GPa$ &&&& \\
\hline
H$_2$& 79.43&0.162& 6.70 &4.837 &1.385 & 1.863\\
He& 291.73&0.224& 7.15 &5.141 &1.391 & 2.009\\
H$_2$O&998.0& 2.20& 7.13 &5.248 &1.359 & 1.882\\
MgO&3580.0& 157.0&4.37&2.465 & 1.772 & 1.904\\
SiO$_2$&4287.0&305.0&4.75&2.983 & 1.592 & 1.767\\
Fe & 8300.0 & 165.0& 5.15&3.080& 1.672 &2.070\\
  \hline
\end{tabular}
\caption{$\rho_0, B_0, n_0$ are chosen to be
a rough consensus of the experimental values at vacuum pressure.
The values of $A_0, A_1, A_2$ follow the prescription of
Eq.~(\ref{par1}) which is slightly adjusted on each iteration of
the secant method searching for the critical density.
First order approximate values are
$A_2=1.95$,$A_1 = \frac{n_0}{n_0-1.95}$ and $A_0 = n_0 - A_2$ where
$n_0 = {\frac{dB}{dP}}_0= B_0^\prime.$}
\label{T1}
\end{table}

The parameters, $\rho_0,B_0$, and $n_0=\frac{dB}{dP}$, were not adjusted and were
set to values which best reflect the 
experimental results~\citep{Zha93,Khai08,Xian10,Duffy95,Panero03,Driver10} 
the rest of these parameters 
are derived from these 
three experimental values. To first approximation these 
parameters are $A_2=1.95$,$A_1 = \frac{n_0}{n_0-1.95}$
(or adjustable for best fit) and $A_0 = n_0 - A_2$ if one
needs pressures only up to the Tera-Pascal range.

At higher pressures we switch to a Thomas-Fermi-Dirac scheme of 
Ref.~\citep{Salpeter:1967zz}. To make this transformation
we find a critical density where we match $B$ and $\frac{dB}{dP}=n$, we then create 
a match with pressure by adding a constant $P_0$ to the Thomas-Fermi-Dirac scheme.
We first assume an initial guess using a 
secant method to find the critical density. The density 
is than iterated until it meets the requirement $B_{TFD}(\rho_c)-B_{classical}(\rho_c)=0$.
At this critical density the derivative of the 
bulk modulus with respect to pressure also matches thanks to the relationship 
in Eq.~(\ref{eq1.4}).
At each iteration of the search we adjust $A_2$ and then we are 
able to set $A_0$ and $A_1$:
\begin{eqnarray}
A_2 &=& n_{TFD}-(n_0-n_{TFD})*(\rho_0/\rho_c)^{\frac{n_0}{n_0-n_{TFD}}}\nonumber\\
A_1 &=& \frac{n_0}{n_0-A_2}\nonumber\\
A_0 &=& n_0 - A_2,\label{par1}
\end{eqnarray}
where $n_{TFD}$ is the past iterative guess using the present 
iteration of $\rho_c$ in Eq.~(\ref{tfd1}) while $n_0$ is set to the rough experimental
consensus value.
The results of the secant method with additional useful parameters 
(atomic number, $A$, and proton number, $Z$) are
in Table~\ref{T2}.

\begin{table}
\centering
\begin{tabular}{ |l || c |c|c|c|c|c|c| }
  \hline
material &$\rho_c$&$B_c$&$n_c$& $P_c$ & $P_0$  & $A$ &$Z$\\
&   $kg/m^3$ & $GPa$ &&$GPa$&$GPa$&& \\
\hline
  H$_2$& $1.690E4$ & $1.154E5$ & 1.866 &$6.165E4$ & $2.782E3$ &2&2 \\
  He & $1.229E4$ & $1.623E4$ & 2.037 &$7.801E3$ & $7.806E2$ &4&2 \\
  H$_2$O&$3.758E5$ & $7.360E6$ & 1.883& $3.900E6$ & $2.027E5$ &18&10\\
  MgO&$1.262E6$ & $4.478E7$ & 1.905& $2.351E7$ & $1.434E6$ &40&20 \\
 SiO$_2$&$1.577E7$ & $3.967E9$ & 1.767& $2.245E9$ & $4.396E7$ &60&30\\
 Fe & $9.736E5$ & $1.998E7$ & 2.071 &$9.631E6$ & $1.2806E6$ &55.85&26\\
  \hline
\end{tabular}
\caption{The critical values
are found by solving for the critical density, $\rho_c$,
using the secant method when the
bulk modulus of our EOS, Eq.~(\ref{eos5}) matches the
bulk modulus of the degenerative
Fermi gas bulk modulus of Thomas-Fermi-Dirac, Eq.~(\ref{tfd5}).
The parameters $P_c$ and $P_0$ are found by forcing
the pressure formula of Eq.~(\ref{tfd5}) to equal the
pressure formula listed in Eq.~(\ref{eos2}) by choosing
an appropriate $P_0$}
\label{T2}
\end{table}
Again no adjustable parameters are in this table, 
they were found using the secant method by requiring that 
the bulk modulus matches at the boundary between the low 
pressure and high pressure theories.

\section{The polytrope index for various equations of state}\label{CC}
Our Equation of state is developed from 
a functional form of $\frac{dB}{dP}$ which is detailed in Appendix \ref{oureos}.
We repeat them here:
\begin{eqnarray}
B^\prime \equiv n &=& \frac{dB}{dP}=A_0\left(\frac{\rho_0}{\rho}\right)^{A_1} + A_2 \\
B &=& B_0e^{\frac{A_0}{A_1}(1-(\frac{\rho_0}{\rho})^{A_1})}\left(\frac{\rho}{\rho_0}\right)^{A_2} \\
P&=& \frac{B_0e^{\frac{A_0}{A_1}}}{A_1}
\left [\left(\frac{\rho}{\rho_0}\right)^{A_2}E_n\left(\frac{A_1+A_2}{A_1}
,\frac{A_0}{A_1}\left(\frac{\rho_0}{\rho}\right)^{A_1}\right)  
- E_n\left(\frac{A_1+A_2}{A_1},\frac{A_0}{A_1}\right)\right ] 
\end{eqnarray}
The equation of state of Stacey also 
starts with $B^\prime=\frac{dB}{dP}$,
\begin{equation}
\frac{1}{B^\prime} = \frac{1}{B^\prime_0}
+\left (1-\frac{B^\prime_\infty}{B^\prime_0}\right ) \frac{P}{B}
\end{equation}
which then one can derive a formula for the bulk modulus and
the density ratio
\begin{eqnarray}
B &=& B_0{\left (1 - B^\prime_\infty\frac{P}{B}\right )}
^{-\frac{B^\prime_0}{B^\prime_\infty}} \\
\ln\frac{\rho}{\rho_0} &=& - \frac{B^\prime_0}{{B^\prime_\infty}^2}
\ln {\left (1 - B^\prime_\infty\frac{P}{B}\right )}
- \left (\frac{B^\prime_0}{{B^\prime_\infty}}-1\right )\frac{P}{B}
\end{eqnarray}

Most equations of state start from a definition of pressure,
one can calculate $n=\frac{dB}{dP}$ from $P(\rho)$ analytically
\begin{equation}
\frac{dB}{dP} = \frac{d}{dP}(\rho\frac{dP}{d\rho})= \rho\frac{dB/d\rho}{B}
\end{equation}

Many of the next equations use $\eta = \frac{\rho}{\rho_0}=\frac{V_0}{V}$.
The Vinet EOS~\citep{JGRB:JGRB6242,0953-8984-1-11-002,earth} is
\begin{eqnarray}
P &=& 3B_0\eta^\frac{2}{3}\left (1-\eta^{-\frac{1}{3}} \right ) \exp(\frac{3}{2}(B_0^\prime-1)(1-\eta^{-\frac{1}{3}})) \\
B &=& 3B_0 \left (\frac{2}{3}\eta^\frac{2}{3}-\frac{1}{3}\eta^\frac{1}{3}+\frac{1}{2}(B_0^\prime-1)(\eta^\frac{1}{3}-1)\right )
\exp(\frac{3}{2}(B_0^\prime-1)(1-\eta^{-\frac{1}{3}})) \\
B^\prime &=& \frac{(\frac{2}{3})^2\eta^\frac{2}{3}-(\frac{1}{3})^2\eta^\frac{1}{3}+\frac{1}{2}(B_0^\prime-1)(\eta^\frac{1}{3}-\frac{1}{3})
+\frac{1}{4}(B_0^\prime-1)^2(1-\eta^{-\frac{1}{3}})}
{\frac{2}{3}\eta^\frac{2}{3}-\frac{1}{3}\eta^\frac{1}{3}+\frac{1}{2}(B_0^\prime-1)(\eta^\frac{1}{3}-1)}
\end{eqnarray}

The Birch-Murnaghan third order EOS is~\citep{Birch,earth}
\begin{eqnarray}
P &=& \frac{3}{2}B_0(\eta^\frac{7}{3}-\eta^\frac{5}{3})
\left (1+\frac{3}{4}(B_0^\prime-4)(\eta^\frac{2}{3}-1)\right ) \\
B &=& \frac{3}{2}B_0\left (\frac{7}{3}\eta^\frac{7}{3}-\frac{5}{3}\eta^\frac{5}{3}+
\frac{3}{4}(B_0^\prime-4)\left ( \frac{9}{3}\eta^\frac{9}{3} 
-2(\frac{7}{3})\eta^\frac{7}{3}+\frac{5}{3}\eta^\frac{5}{3}\right )\right ) \\
B^\prime &=& \frac{(\frac{7}{3})^2\eta^\frac{7}{3}-(\frac{5}{3})^2\eta^\frac{5}{3}
+\frac{3}{4}(B_0^\prime-4)\left ( (\frac{9}{3})^2\eta^\frac{9}{3}
-2(\frac{7}{3})^2\eta^\frac{7}{3}+(\frac{5}{3})^2\eta^\frac{5}{3}\right )}
{\frac{7}{3}\eta^\frac{7}{3}-\frac{5}{3}\eta^\frac{5}{3}+
\frac{3}{4}(B_0^\prime-4)\left ( \frac{9}{3}\eta^\frac{9}{3}
-2(\frac{7}{3})\eta^\frac{7}{3}+\frac{5}{3}\eta^\frac{5}{3}\right )}
\end{eqnarray}

One must be careful with the Roy-Roy equation of state, they have
two different forms. The one they consider
superior~\citep{0953-8984-11-50-328,0953-8984-18-46-015,0953-8984-17-39-007},
which has a logarithmic form, does not have a physical  $B^\prime_\infty$
asymptote (it is designed to go no higher than $10^{12}$ Pa). The one we
depict in Fig.~\ref{fig6} and Appendix~\ref{CC} has
a physical asymptotic form and
has $B^\prime_\infty=5/3$~\citep{PSSB:PSSB125,0953-8984-11-50-328}.

The ~\cite{0953-8984-11-50-328} EOS is
\begin{eqnarray}
\eta &=& 1+aP(1+bP)^c \\
B &=& \frac{aP+abP^2+(1+bP)^{1-c}}
{a+ab(1+c)P} \\
B^\prime &=& \frac{abc(1+bP)^{-c}(-b(1+c)P-2)+a^2(1+bP(b(1+c)P+2))}
{(a+ab(1+c)P)^2}
\end{eqnarray}

where
\begin{eqnarray}
a &=& \frac{1}{B_0} \\
b &=& \frac{1+2B_0^\prime-5(B_0^\prime)^2-1}
{6B_0(1-B_0^\prime)} \\
c &=& \frac{(3(1-B_0^\prime)^2)}{2B_0^\prime-5(B_0^\prime)^2-1}
\end{eqnarray}

The DFT EOS equation of state~\citep{Geng12} is
\begin{eqnarray}
P &=& 10^{N(\rho)} \\
B &=& -\frac{10^{N(\rho)}
(a_1\rho^{4/3}+2a_2\rho+3a_3\rho^{2/3}+4a_4\rho^{1/3}+5a_5)\ln(10)}{3\rho^{5/3}} \\
B^\prime &=& \frac{-25a_5^2\ln(10)-25a_5\rho^{5/3}+40a_4a_5\rho^{1/3}\ln(10)+30a_3a_5\rho^{2/3}\ln(10)}
{15a_5\rho^{5/3}+12a_4\rho^{6/3}+9a_3\rho^{7/3}+6a_2\rho^{8/3}+3a_1\rho^{9/3}} \nonumber\\
&+&\frac{20a_2a_5\rho\ln(10)+5a_1a_5\rho^{4/3}\ln(100)}
{15a_5\rho^{5/3}+12a_4\rho^{6/3}+9a_3\rho^{7/3}+6a_2\rho^{8/3}+3a_1\rho^{9/3}} \nonumber\\
&-&\frac{16a_4^2\rho^{2/3}\ln(10)+9a_3\rho^{7/3}+4a_2\rho^{8/3}+a_1\rho^{9/3}+\rho^{4/3}(3a_3+2a_2\rho^{1/3}+a_1\rho^{2/3})^2\ln(10)}
{15a_5\rho^{5/3}+12a_4\rho^{6/3}+9a_3\rho^{7/3}+6a_2\rho^{8/3}+3a_1\rho^{9/3}} \nonumber\\
&-&\frac{16a_4\rho^{6/3}+24a_3a_4\rho^{3/3}\ln(10)+8a_4a_1\rho^{5/3}\ln(10)+8a_4a_2\rho^{4/3}\ln(100)}
{15a_5\rho^{5/3}+12a_4\rho^{6/3}+9a_3\rho^{7/3}+6a_2\rho^{8/3}+3a_1\rho^{9/3}}
\end{eqnarray}
with $a_0 = 1.0683, a_1 = 19.1824(2696^{1/3}), a_2 = -36.3776(2696^{2/3}), a_3 = 28.5165(2696^{3/3}),a_4 = -10.6068(2696^{4/3}),
a_5 = 1.5224(2696^{5/3})$.
These complicated functions were calculated using {\it Mathematica}. The power functions are 
\begin{equation}
N(\rho) = 9+a_0+\frac{a_1}{\rho^{1/3}}+\frac{a_2}{\rho^{2/3}}+\frac{a_3}{\rho^{3/3}}+\frac{a_4}{\rho^{4/3}}+\frac{a_5}{\rho^{5/3}} \\
\end{equation}
In Fig.~6 we also attempted to analyze experimental data independent of a universal EOS.
We take high precision pressure-density data that enters the $10^{9}$Pa range for light elements
and the $10^{10}$Pa range for heavier elements. It is also advantageous to have at
least two points of low pressure data ($<10^8$Pa) that anchors the fit towards the correct
origin. In the case of the hydrogen data we found that a 5th order polynomial gave a 
very low chi-squared. Once we have a polynomial for $P$, calculating the bulk modulus 
and the polytrope index are trivial.
\begin{eqnarray}
P(GPa) &=& a_0+a_1\rho+a_2\rho^2+a_3\rho^3+a_4\rho^4+a_5\rho^5\\
B(GPa) &=& a_1\rho+2a_2\rho^2+3a_3\rho^3+4a_4\rho^4+5a_5\rho^5\\
n &=& \frac{a_1\rho+4a_2\rho^2+9a_3\rho^3+16a_4\rho^4+25a_5\rho^5}
{a_1\rho+2a_2\rho^2+3a_3\rho^3+4a_4\rho^4+5a_5\rho^5}
\end{eqnarray}
with $a_0 = 2.26603 GPa$, $a_1 = -4.52911\times 10^{-2} GPa/(kg/m^3)^1$,
$a_2 = 2.16832\times 10^{-4}GPa/(kg/m^3)^2$, $a_3 = 1.77254\times 10^{-8} GPa/(kg/m^3)^3$,
$a_4=1.64780\times 10^{-10}GPa/(kg/m^3)^4$, $a_5 = -1.48451\times 10^{-13}GPa/(kg/m^3)^5$.
It also helps to normalize the fitting routine, have as your dependent fitting 
variable $\rho/(2\rho_{max})$ so that your domain is much less than one, the
range should also be small, here we fit it to Giga Pascals.

\end{document}